\documentclass[reprint,amsmath,amssymb,aps]{revtex4-1}
\usepackage{graphicx}% Include figure files
\usepackage{dcolumn}% Align table columns on decimal point
\usepackage{bm}% bold math
\usepackage{hyperref}
\hypersetup{pdftex,colorlinks=true,allcolors=blue,bookmarksopen=true}
\newcommand\myFigureWidth{0.21}

\begin{document}
\title{Correlation and Relaxation Times for a Stochastic Process with a Fat-Tailed Steady-State Distribution}

\author{Z. Liu}
\author{R. A. Serota} 
 \email{serota@ucmail.uc.edu}
\affiliation{ 
  Department of Physics, University of Cincinnati, Cincinnati, Ohio, 45221-0011
}

\date{\today}

\begin{abstract}
We study a stochastic process defined by the interaction strength for the return to
the mean and a stochastic term proportional to the magnitude of the variable. Its
steady-state distribution is the Inverse Gamma distribution, whose power-law tail
exponent is determined by the ratio of the interaction strength to stochasticity. Its
time-dependence is characterized by a set of discrete times describing relaxation of
respective cumulants to their steady-state values. We show that as the progressively
lower cumulants diverge with the increase of stochasticity, so do their relaxation
times. We analytically  evaluate the correlation function and show that it is determined by
the longest of these times, namely the inverse interaction strength, which is also the
relaxation time of the mean. We also investigate relaxation of the entire distribution to the steady state and the distribution of relaxations times, which we argue to be Inverse Gaussian.
\end{abstract}

\maketitle

\section{Introduction\label{Introduction}}

For a stochastic process with a steady-state distribution, a natural question arises of
what is the relaxation time towards the stationary process. In other words, when
the initial values of the variables are chosen from a distribution that differs from
that of the steady-state distribution, what is the time for the distribution to settle
into its steady-state and for various quantities to achieve their stationary values.

Of particular interest is the situation when the steady-state distribution has power-law ("fat") tails and, thus, the diverging cumulants -- especially the divergent lowest
cumulants: variance or even mean. In the latter circumstance, one needs to devise
the means of ascertaining, including numerically, that the steady-state distribution
has been reached, especially in the circumstance when the latter may be unknown
analytically.

Another point of interest for such processes is that of the relationship between the
correlation and relaxation times. A related issue is that of relevant time scales and steady-state distributions in strongly (power-law) correlated time series.

In this paper we study a stochastic Ito process 
\begin{equation}
\mathrm{d}x = -J(x-1) \mathrm{d}t + \sqrt{2}\sigma x \mathrm{d}B
\label{correlationPaperItoProcess}
\end{equation}
with the Inverse Gamma (IGa) steady-state distribution, 
\begin{equation}
P_0(x)=\frac{e^{-\frac{J}{x\sigma^2}} \left( \frac{J}{x\sigma^2} \right)^{2 + \frac{J}{\sigma^2}}}{\frac{J}{\sigma^2} \Gamma\left( 1+\frac{J}{\sigma^2} \right)} , \qquad x\ge0
\label{correlationPaperSteadyStateDistribution}
\end{equation}
This process is a limiting case of the Generalized Inverse Gamma (GIGa) process, which describes a stochastic birth-death model, which appears in diverse contexts, such as generalized Bouchaud-M{\'e}zard (BM) network model of economic exchange \cite{bouchaud2000wealth,ma2013distribution}, ontogenetic mass distribution \cite{west2012allometry,holden2013change} and market volatility \cite{ma2014model}. The first term in eq. (\ref{correlationPaperItoProcess}) describes the reversion to the (unit) mean characterized by the interaction strength $J>0$ and $\mathrm{d}B$ in the second term is the Wiener term, with $\sigma>0$ characterizing stochasticity. 

The main time dependence of the cumulants of the time-dependent distribution is given by 
\begin{equation}
\kappa_n(t) \propto \frac{1 - e^{-\lambda _n t}}{\lambda _n}
\label{correlationPaperTimeDependenceCummulants}
\end{equation}
where
\begin{equation}
\lambda _n = n[J - \sigma^2(n - 1)], \qquad 1\le n\le 1+J/\sigma^2
\label{correlationPaperTimeDependenceCummulantsLambda}
\end{equation}
which is predicated, of course, on the assumption that the initial values are not
chosen from the steady-state distribution (in particular, $\kappa_n(0)\ne \kappa_n(\infty)$ for all $n$, the
latter being the $n$th cumulant of the steady-state distribution), since in such
case the process is stationary. As stochasticity increases via $\sigma^2$, progressively lower
cumulants become divergent, $\kappa_n \propto \lambda^{-1}_n$ as $\sigma^2(n-1)\rightarrow J$ for progressively smaller $n$, 
as do their respective relaxation time $\tau^{(n)}_{relax} \sim \lambda^{-1}_{n}$. Once $\sigma^2 > J$, cumulants of the
steady-state distribution no longer exist with the exception of the mean, whose
relaxation time is $\tau^{(1)}_{relax} \sim \lambda^{-1}_{1} \sim J^{-1}$. 

As will be discussed later, for $J>\sigma^2$, the IGa process is characterized by the correlation
function 
\begin{equation}
\langle \delta x(t+\tau)\delta x(t)\rangle = \frac{e^{-J\tau}(1-e^{-2(J-\sigma^2)t})}{\frac{J}{\sigma^2}-1} 
% \xrightarrow{t\gg J^{-1}} \stackrel{t\gg J^{-1}}{\longrightarrow} \frac{e^{-J\tau}}{\frac{J}{\sigma^2} - 1}
\label{correlationPaperIGaCorrelationFunction}
\end{equation}
and becomes divergent as $\sigma^2\ge J$ (conversely, for $J\gg \sigma^2$, when we can set $x\approx 1$ in
the stochastic term, we recover the well-known correlation function of the Ornstein-Uhlenbeck process.). This follows from the eigenvalue analysis of the
Fokker-Planck (FP) equation \cite{stratonovich1963topics, schenzle1979multiplicative,risken1996fokker}. In this formalism, the eigenvalues are given by eq. (\ref{correlationPaperTimeDependenceCummulantsLambda}),
but only half of them correspond to a complete set of orthogonal eigenfunctions that
have $P_0(x)$ as their "attractor", 
\begin{equation}
\lambda_n=n[J-\sigma^2(n-1)], \qquad 1\le n\le \frac{1+J/\sigma^2}{2}
\label{correlationPaperEigenvalues}
\end{equation}

The goal of this paper is to examine the relaxation towards steady-state distribution,
especially on approach to and in the regime when $\sigma^2>J$. In Section \ref{EigenvalueFormalismForStochasticIGaProcess}, we discuss
the analytical eigenvalue formalism and, in particular, the correlation 
function (\ref{correlationPaperIGaCorrelationFunction}). In Section \ref{CumulantRelaxationForStochasticIGaProcess}, we derive and numerically examine the cumulant relaxation.
In Section IV, we study relaxation of the distribution as a whole to IGa and argue that relaxation times generated by (\ref{correlationPaperItoProcess}) along different paths are distributed as Inverse Gaussian (IG). 

\section{Eigenvalue Formalism for Stochastic IG\lowercase{a} Process\label{EigenvalueFormalismForStochasticIGaProcess}}
The FP equation for the stochastic IGa process (\ref{correlationPaperItoProcess}) can be written as 
\begin{equation}
\frac{\partial P(x,\,t)}{\partial t}=J\frac{\partial [(x-1)P(x,\,t)]}{\partial x} + \sigma^2 \frac{\partial^2 [x^2 P(x,\,t)]}{\partial x^2}
\label{correlationPaperFokkerPlanckEquation}
\end{equation}
We seek solution in the standard form \cite{stratonovich1963topics}:
\begin{equation}
P(x,\,t)=P_0(x)+P(\lambda;\,x)e^{-\lambda t}
\label{correlationPaperFokkerPlanckEquationSolutionForm}
\end{equation}
where $P_0(x)$ is given by eq. (\ref{correlationPaperSteadyStateDistribution}) and $P(\lambda;\,x)$, $\lambda>0$, are the solutions of the eigenvalue problem 
\begin{equation}
\left( x^2 P(\lambda;\,x) \right)'' + \frac{J}{\sigma^2}\left( (x-1)P(\lambda;\,x) \right)' + \frac{\lambda}{\sigma^2}P(\lambda;\,x) = 0
\label{correlationPaperFokkerPlanckEquationForPLambda}
\end{equation}
(Obviously, $P_0(x)$ corresponds to $\lambda=0$ in (\ref{correlationPaperFokkerPlanckEquationForPLambda}).) Physically, since $P_0(x)$ describes the stationary state, the time-dependent terms in (\ref{correlationPaperFokkerPlanckEquationSolutionForm}) are the deviations that describe relaxation to the
steady-state distribution and $\lambda^{-1}$ are thus
relaxation times. \footnote {See also papers by P. Jung and coauthors in \cite{prigogine1995advances,talkner1995new}.} This immediately reveals such properties of $P(\lambda;\,x)$ as
$P(\lambda;\,0)=P'(\lambda;\,0)=0$ and zero probability current at $x=0$, $\int^{\infty}_0 P(\lambda;\,x) \mathrm{d}x = 0$ and a
power-law decay of $P(\lambda;\,x)$ at $x\rightarrow \infty$, which also follow from the general formalism
\cite{schenzle1979multiplicative,risken1996fokker} (see below). It also indicates that eigenfunctions $P(\lambda;\,x)$ have to be real (it is obvious that two independent
real solutions of (\ref{correlationPaperFokkerPlanckEquation}) can always be constructed).

Solutions of eq. (\ref{correlationPaperFokkerPlanckEquationForPLambda}) are given by
\begin{widetext}
\begin{equation}
\label{correlationPaperFokkerPlanckEquationSolutions}
% \begin{split}
P_{1,\,2}(\lambda;\,x) \propto 
% & 
\left( \frac{J}{x\sigma^2} \right)^{\frac{3}{2}+\frac{J}{2\sigma^2}\pm \sqrt{\left( \frac{J}{2\sigma^2}+\frac{1}{2} \right)^2 - \frac{\lambda}{\sigma^2}}} 
% \\
% & 
% \qquad 
{}_1F_1 \left( \frac{3}{2}+\frac{J}{2\sigma^2} \pm \sqrt{\left( \frac{J}{2\sigma^2}+\frac{1}{2} \right)^2 - \frac{\lambda}{\sigma^2}},\, 1\pm 2\sqrt{\left( \frac{J}{2\sigma^2}+\frac{1}{2} \right)^2 - \frac{\lambda}{\sigma^2}},\, -\frac{J}{x\sigma^2} \right)
% \end{split}
\end{equation}
\end{widetext}
where ${}_1F_1$ is the Kummer's confluent hypergeometric function and so 
\begin{eqnarray}
% \begin{equation}
P_{1,\,2}(\lambda;\,x) \xrightarrow{x\rightarrow 0} 
&& 
\frac{\Gamma\left( 1\pm 2 \sqrt{\left( \frac{J}{2\sigma^2}+\frac{1}{2} \right)^2 - \frac{\lambda}{\sigma^2}} \right)}{\Gamma\left( -\left( \frac{J}{2\sigma^2}+\frac{1}{2} \right)\pm \sqrt{\left( \frac{J}{2\sigma^2}+\frac{1}{2} \right)^2 - \frac{\lambda}{\sigma^2}} \right)} 
\nonumber \\ && 
+ \mathrm{const}\cdot e^{-\frac{J}{x\sigma^2}} x^{-\left( 2+\frac{J}{\sigma^2} \right)}
% \end{equation}
\end{eqnarray}
where $\Gamma$ is the Gamma function. The above restriction on $P(\lambda;\,x)$ requires that the argument of the Gamma function in the denominator is negative integer or zero, which results in the discrete spectrum given by eq. (\ref{correlationPaperTimeDependenceCummulantsLambda}) and the respective functions given by 
\begin{eqnarray}
% \begin{widetext}
% \begin{equation}
&& P_{1,\,2}(\lambda_n;\,x) 
\equiv P_n(x) 
% \nonumber 
\\ && 
\propto \left( \frac{J}{x\sigma^2} \right)^{2-n+\frac{J}{\sigma^2}} {}_1F_1 \left( 2-n+\frac{J}{\sigma^2}, \, 2-2n+\frac{J}{\sigma^2}, \, -\frac{J}{x\sigma^2} \right) 
\nonumber
% \end{equation}
% \end{widetext}
\end{eqnarray}
under the condition that $J\ge (2n-1)\sigma^2$, that is eq. (\ref{correlationPaperEigenvalues}), for $P_1$ and $J\le(2n-1)\sigma^2$ 
for $P_2$ respectively.

Notice that $\lambda_n$ has a maximum as a function of $n$ 
\begin{equation}
\lambda_{max} = \lambda_{n_{max}} = \frac{\sigma^2}{4}\left( \frac{J}{\sigma^2}+1 \right)^2, \, n_{max}=\frac{1}{2}\left( 1+\frac{J}{\sigma^2} \right)
\label{correlationPaperMaximumLambda}
\end{equation}
so that $P_1$ and $P_2$ correspond to the two branches of parabola (\ref{correlationPaperTimeDependenceCummulantsLambda}) that defines $\lambda_n$, to
the left and to the right of the maximum respectively and have identical properties
under transformation $n\leftrightarrow 1+J/\sigma^2-n$. Functions $P_1(\lambda_n;\,x)$, with $1\le n\le n_{max}$ 
(that is (\ref{correlationPaperEigenvalues})), form a complete set of discrete orthogonal eigenfunctions [6,7] that
correspond to $P_0(x)$. (Notice that under this constraint, the argument of the Gamma
function in the numerator is positive for $P_1(\lambda_n;\,x)$.) It is also clear from (\ref{correlationPaperMaximumLambda}) that
discrete spectrum corresponds to the positive argument of the square root in (\ref{correlationPaperFokkerPlanckEquationSolutions})
and that conversely, for $\lambda>\lambda_{max}$, the spectrum is continuous. In the latter case,
$P_1(\lambda;\,x)=P^*_2(\lambda;\,x)$ and the two real, independent solutions of (\ref{correlationPaperFokkerPlanckEquationForPLambda}) are, respectively,
$\Re[P_1(\lambda;\,x)]$ and $\Im[P_1(\lambda;\,x)]$. The linear combination 
\begin{eqnarray}
% \begin{equation}
P_{cont}(\lambda;\,x) \propto 
&& 
\Re[P_1(\lambda;\,x)]\Im[P_1(\lambda;\,0)] 
\nonumber \\ && 
- \Re[P_1(\lambda;\,0])]\Im[P_1(\lambda;\,x)]
% \end{equation}
\end{eqnarray}
is then formed to satisfy the boundary conditions at $x=0$ for functions (\ref{correlationPaperFokkerPlanckEquationSolutions}), namely that $P_{cont}(\lambda;\,0) = P'_{cont}(\lambda;\,0) = 0$. Both continuous and discrete spectrum function decay as
\begin{equation}
P(\lambda;\,x) \propto e^{-\frac{J}{x\sigma^2}}x^{-\left( 2+\frac{J}{\sigma^2} \right)}, \qquad x\rightarrow 0
\end{equation}
while
\begin{equation}
P_n(x) \propto \left( \frac{J}{x\sigma^2} \right)^{2-n+\frac{J}{\sigma^2}}, \qquad x\rightarrow \infty
\label{correlationPaperPLimitInfinity}
\end{equation}
for discrete and 
\begin{equation}
P_{cont}(\lambda;\,x)\propto \left( \frac{J}{x\sigma^2} \right)^{\frac{3}{2} + \frac{J}{2\sigma^2}}, \qquad x\rightarrow \infty
\end{equation}
for continuous spectrum respectively.

Our results are consistent with the general theory of eigenfunction expansion \cite{schenzle1979multiplicative,risken1996fokker},
for the IGa process. For instance, in notations of \cite{risken1996fokker}, the form of potential
\begin{equation}
\Phi(x) = -\ln P_0(x) = \frac{J}{x\sigma^2} + \left( 2+\frac{J}{\sigma^2} \right) \ln x
\end{equation}
confirms the aforementioned property that $P(\lambda;\,0) = P'(\lambda;\,0) = 0$ and zero current
at $x=0$. In terms of the transformation of the FP into a one-dimensional
Schr$\mathrm{\ddot{o}}$dinger equation \cite{risken1996fokker}, the potential for the latter is Kepler-like
\begin{equation}
V(x) = \lambda_{max} - \frac{J}{2x}\left( 1+\frac{J}{\sigma^2} \right) + \frac{J}{4x^2} \left( 2+\frac{J}{\sigma^2} \right)
\end{equation}
where $\lambda_{max}$ is given by (\ref{correlationPaperMaximumLambda}). From (\ref{correlationPaperPLimitInfinity}) it immediately follows, for instance, that
since $V(x)>0$ all energy eigenvalues $\lambda$ are positive and that the energy spectrum is
discrete for $\lambda<\lambda_{max}$ and continuous for $\lambda>\lambda_{max}$.

In this formalism \cite{schenzle1979multiplicative,risken1996fokker}, the correlation function in the steady state is given by 
\begin{equation}
\langle\delta x(t+\tau)\delta x(t)\rangle = \sum_n g^2_n e^{-\lambda_n \tau} + \int g(\lambda)e^{-\lambda \tau} \mathrm{d}\lambda
\end{equation}
where
\begin{equation}
g = \int xP(\lambda;\,x)\mathrm{d}x = \int \delta x P(\lambda;\,x)\mathrm{d}x
\end{equation}
where $P(\lambda;\,x)$ are the properly normalized eigenfunction \cite{schenzle1979multiplicative,risken1996fokker}. It turns out that for
the IGa process all $g$'s are zero except one, $g_1$, for $n=1$ in (\ref{correlationPaperEigenvalues}). For $J>\sigma^2$ we find 
\begin{equation}
P_1(\lambda_1;\,x) = \frac{e^{-\frac{J}{x\sigma^2}}(x-1)\left( \frac{J}{x\sigma^2} \right)^{2+\frac{J}{\sigma^2}} }{\left( \frac{J}{\sigma^2} \right)^2 \sqrt{\Gamma\left( \frac{J}{\sigma^2} \right) \Gamma\left( \frac{J}{\sigma^2} - 1 \right) } }
\end{equation}
such that
\begin{equation}
\int^\infty_0 \frac{P^2_1(\lambda_1;\,x)}{P_0(x)} \mathrm{d}x = 1
\end{equation}
and 
\begin{equation}
g_1 = \int^\infty_0 P_1(\lambda_1;\,x)x\mathrm{d}x =\frac{1}{\sqrt{{\frac{J}{\sigma^2} -1}}}
\end{equation}
so that the correlation function in the steady state is given by 
\begin{equation}
\langle \delta x(t+\tau)\delta x(t) \rangle = \frac{e^{-J\tau}}{\frac{J}{\sigma^2} - 1}
\label{correlationPaperCorrelationFunctionInSteadyState}
\end{equation}
which is just a $t\gg (J-\sigma^2)^{-1}$ limit of (\ref{correlationPaperIGaCorrelationFunction}); at $\tau=0$ we recover the variance of the
IGa distribution and for $J\gg\sigma^2$ the Ornstein-Uhlenbeck result. 
\footnote{In this formalism, the spectrum of the OU process is discrete, $\lambda_n= Jn$, with
$g_n=g_1 \delta_{n1} = \sqrt{\sigma^2/J} \delta_{n1}$.}
The correlation function diverges as $\sigma^2\rightarrow J$ and does not exist for $J<\sigma^2$ -- a direct result of the
heavy tail of the probability distribution function. 
\footnote{An attempt to carry over the results obtained in \cite{schenzle1979multiplicative} for the correlation function of a
particular stochastic process to the correlation function of the IGa process was made
in \cite{bouchaud2000wealth}. However, since it used the transformation of variable that involved a negative
power, the evaluated quantity was actually not the correlation function.}
Notice that the normalized
correlation function
\begin{equation}
\frac{\langle \delta x(t+\tau)\delta x(t) \rangle}{\langle \delta x(t)^2 \rangle} = e^{-J\tau}
\end{equation}
is the same as for Ornstein-Uhlenbeck process and formally exists even for $J<\sigma^2$;
numerically, of course, the variance is always finite and the normalized correlation
function can be, in principle, calculated in the latter regime (with obvious caveats).

\begin{figure}[!htbp]
\centering
\begin{tabular}{cc}
\includegraphics[width = \myFigureWidth \textwidth]{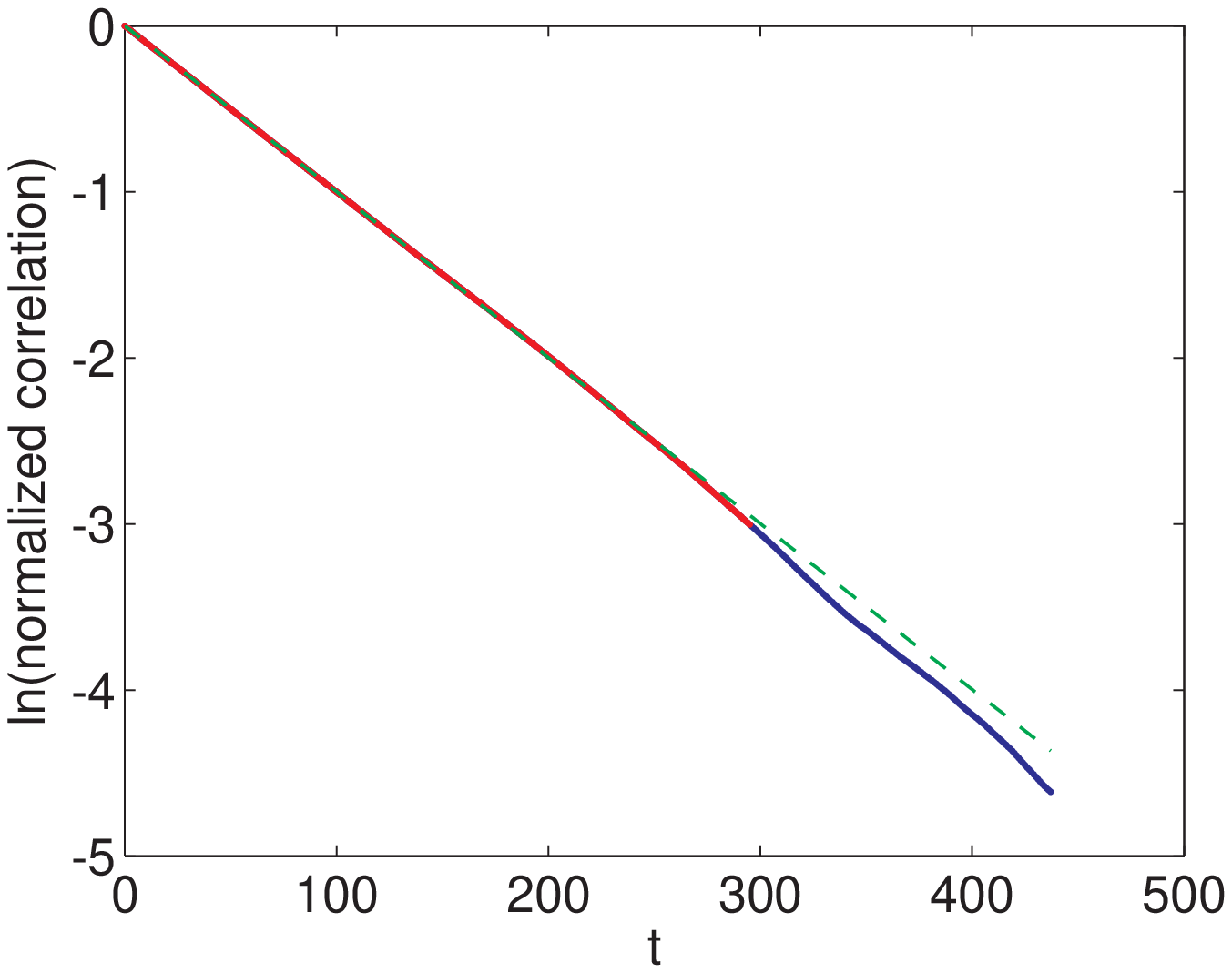} &
\includegraphics[width = \myFigureWidth \textwidth]{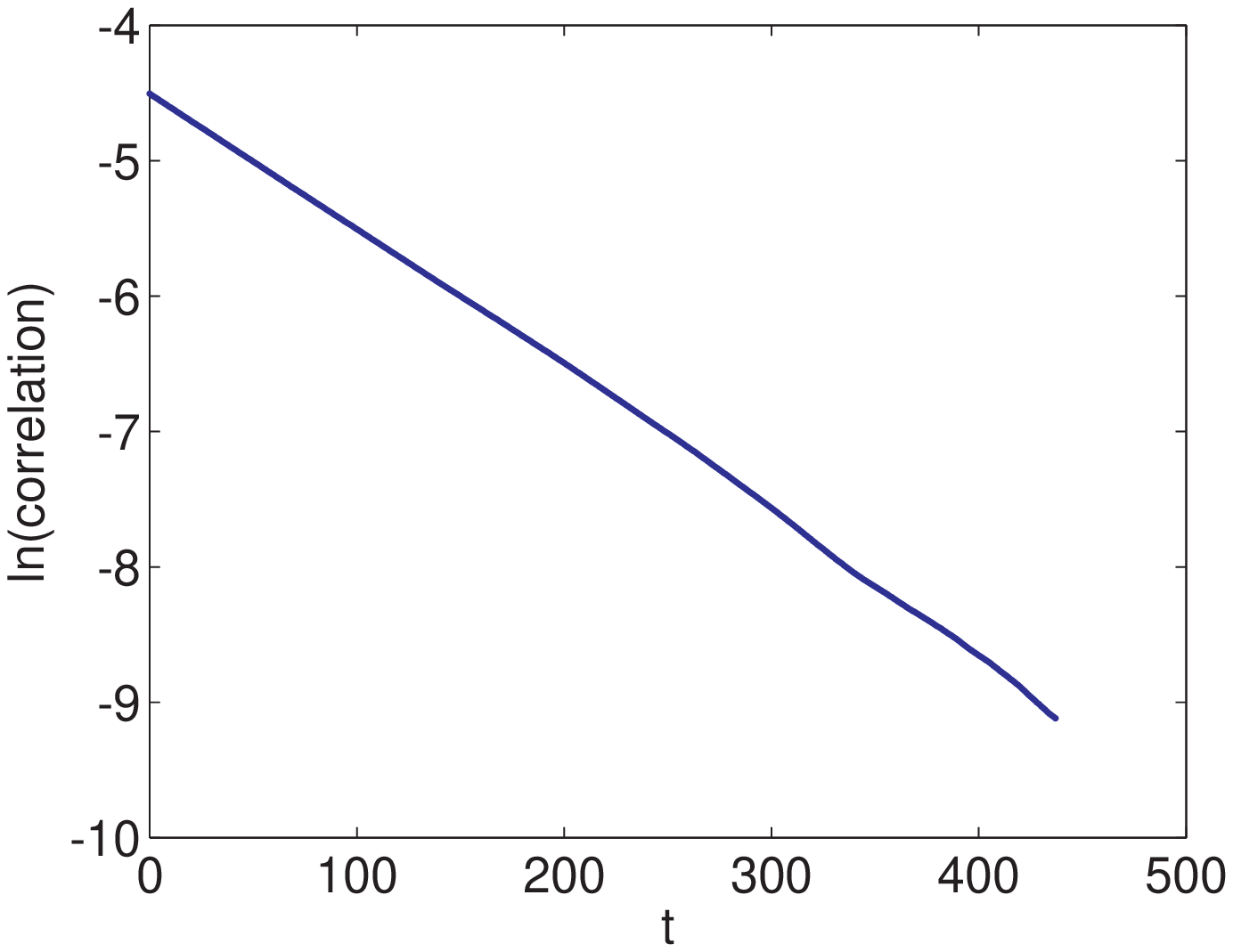} \\
\includegraphics[width = \myFigureWidth \textwidth]{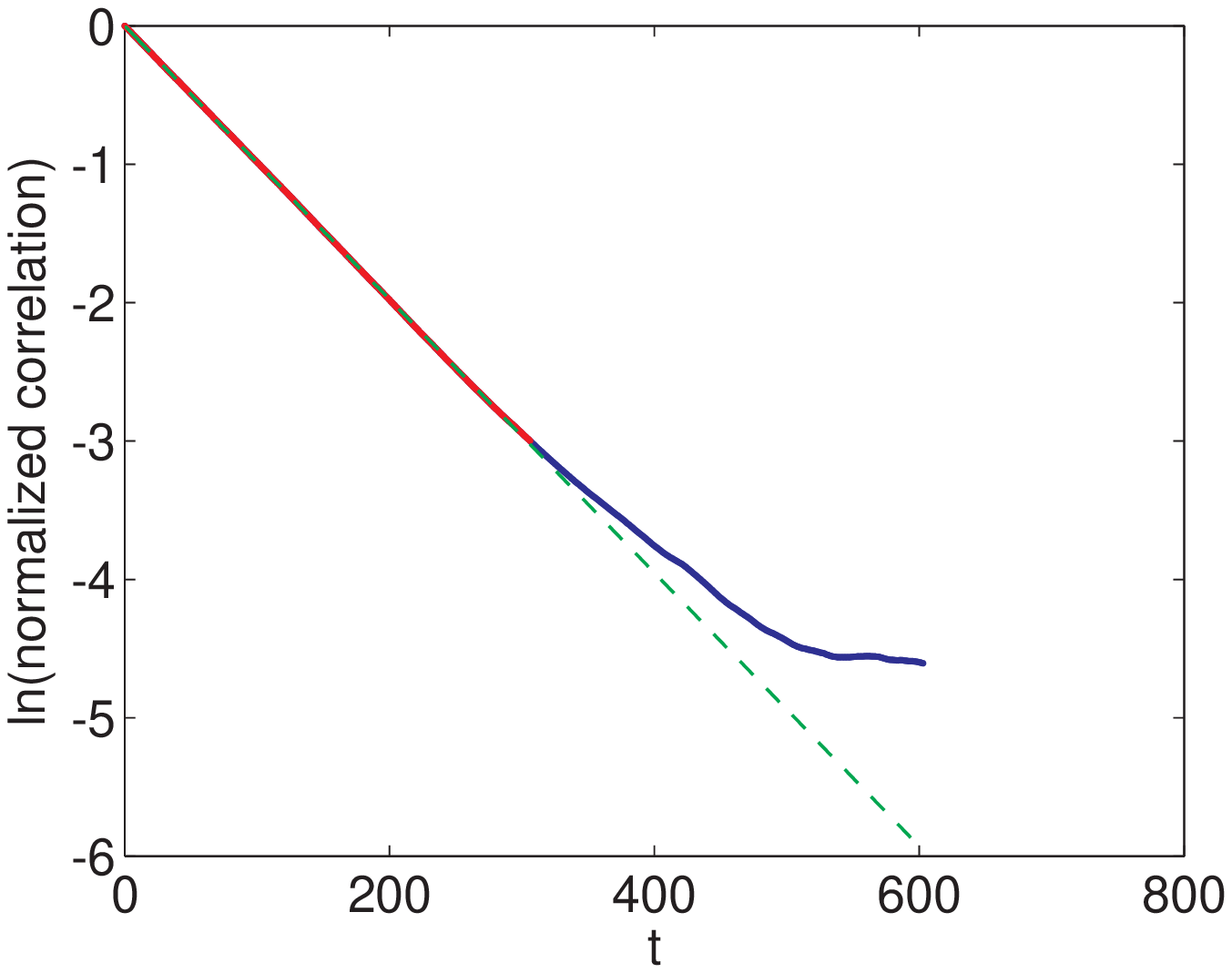} &
\includegraphics[width = \myFigureWidth \textwidth]{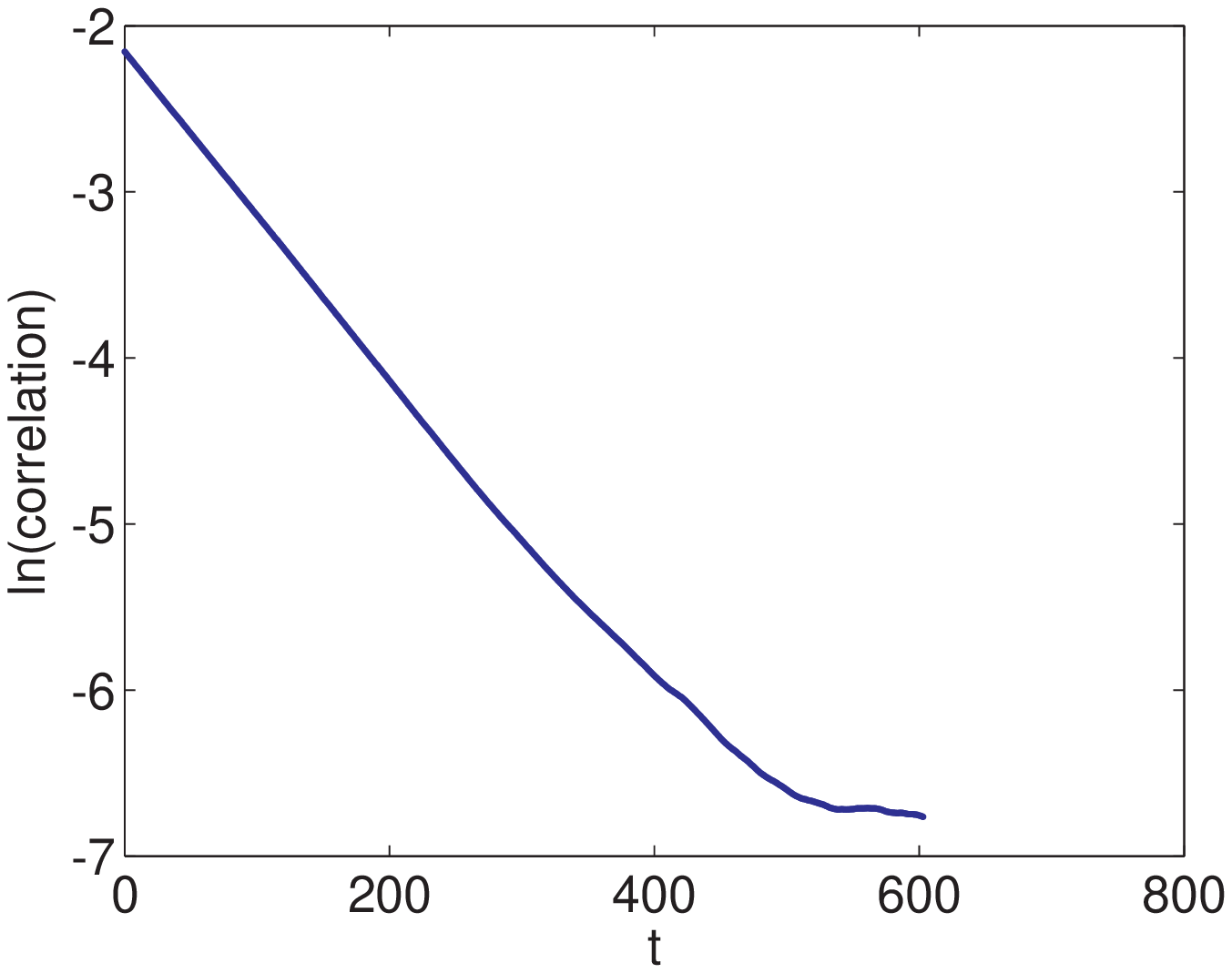} \\
\includegraphics[width = \myFigureWidth \textwidth]{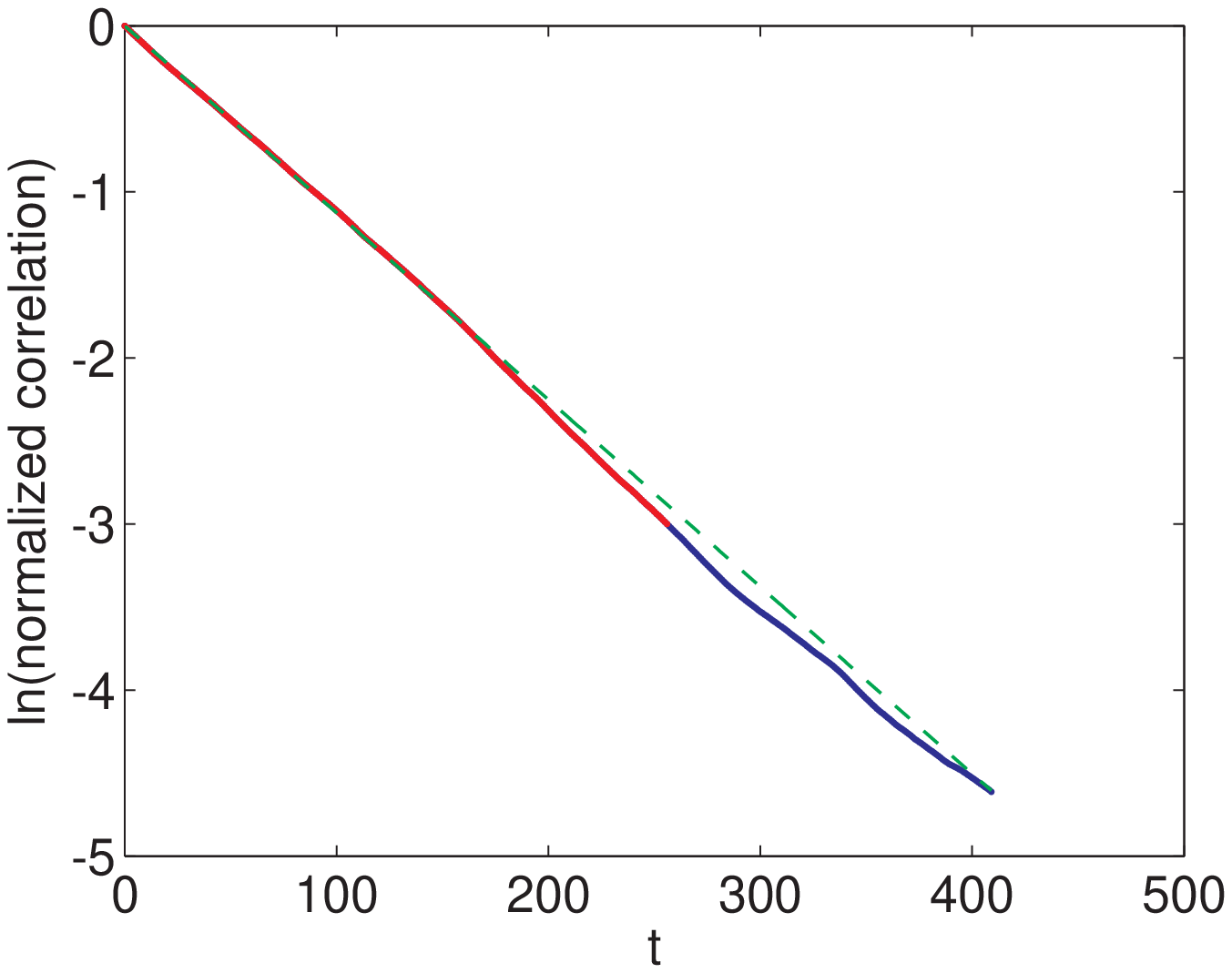} &
\includegraphics[width = \myFigureWidth \textwidth]{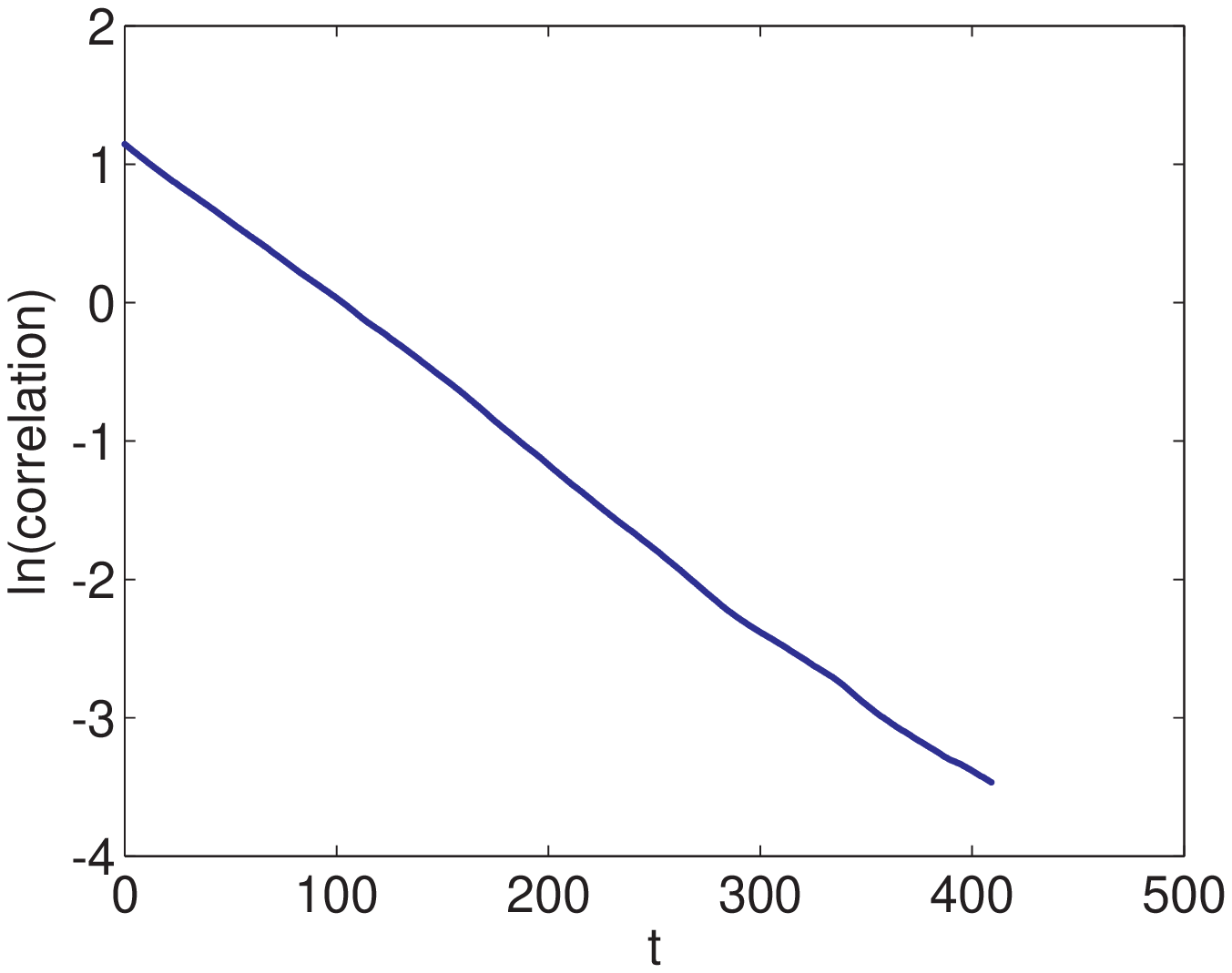} \\
\includegraphics[width = \myFigureWidth \textwidth]{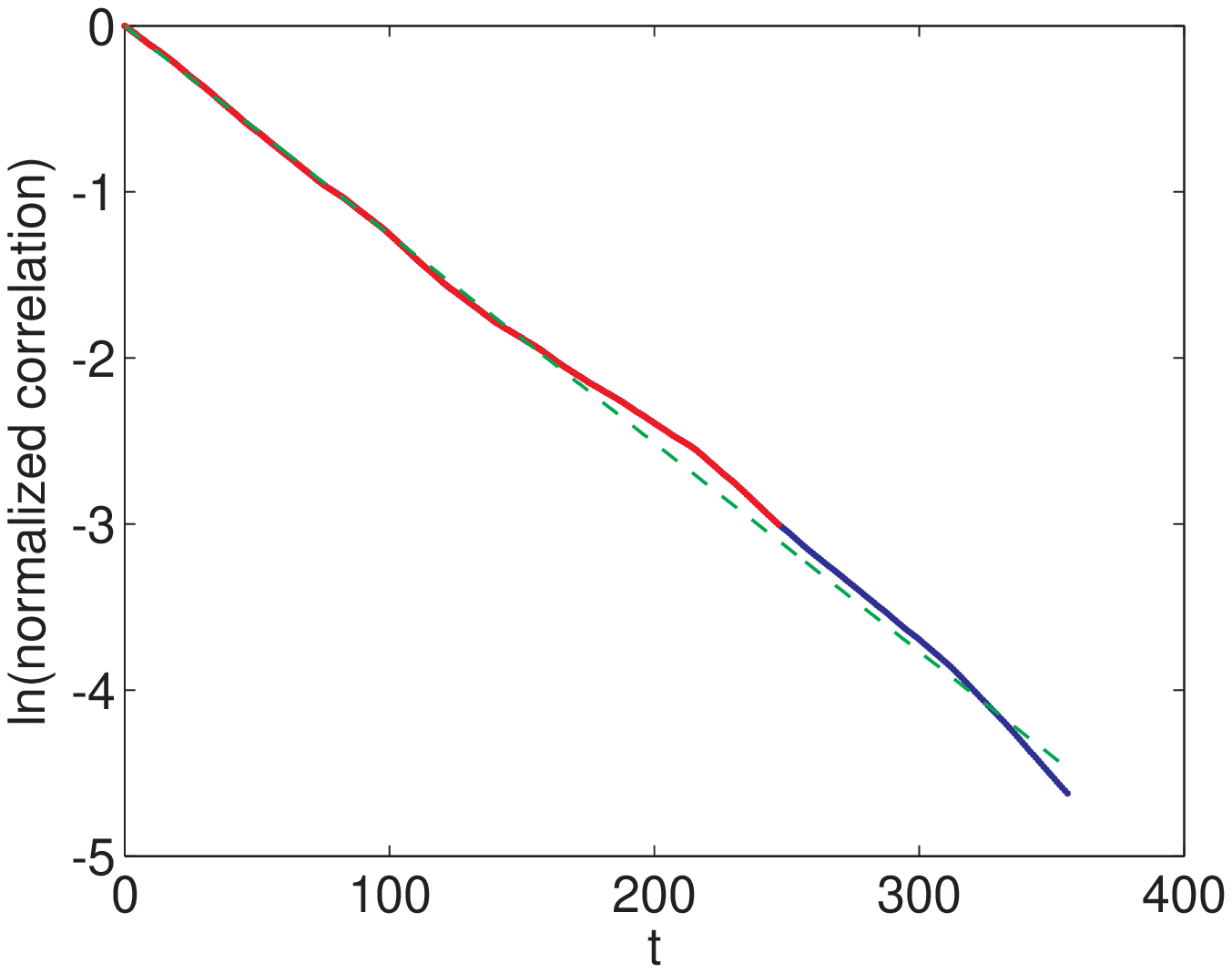} &
\includegraphics[width = \myFigureWidth \textwidth]{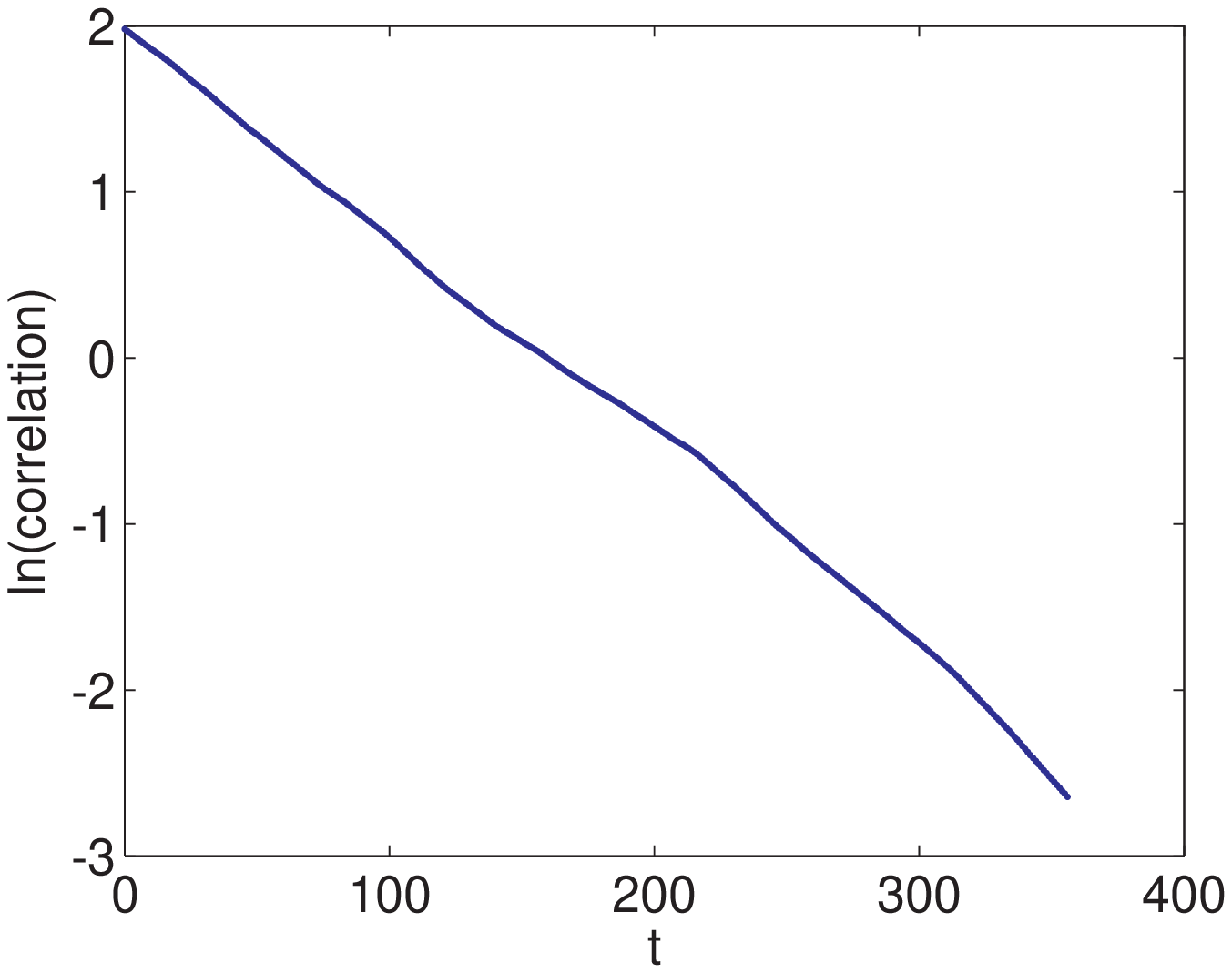} \\
\includegraphics[width = \myFigureWidth \textwidth]{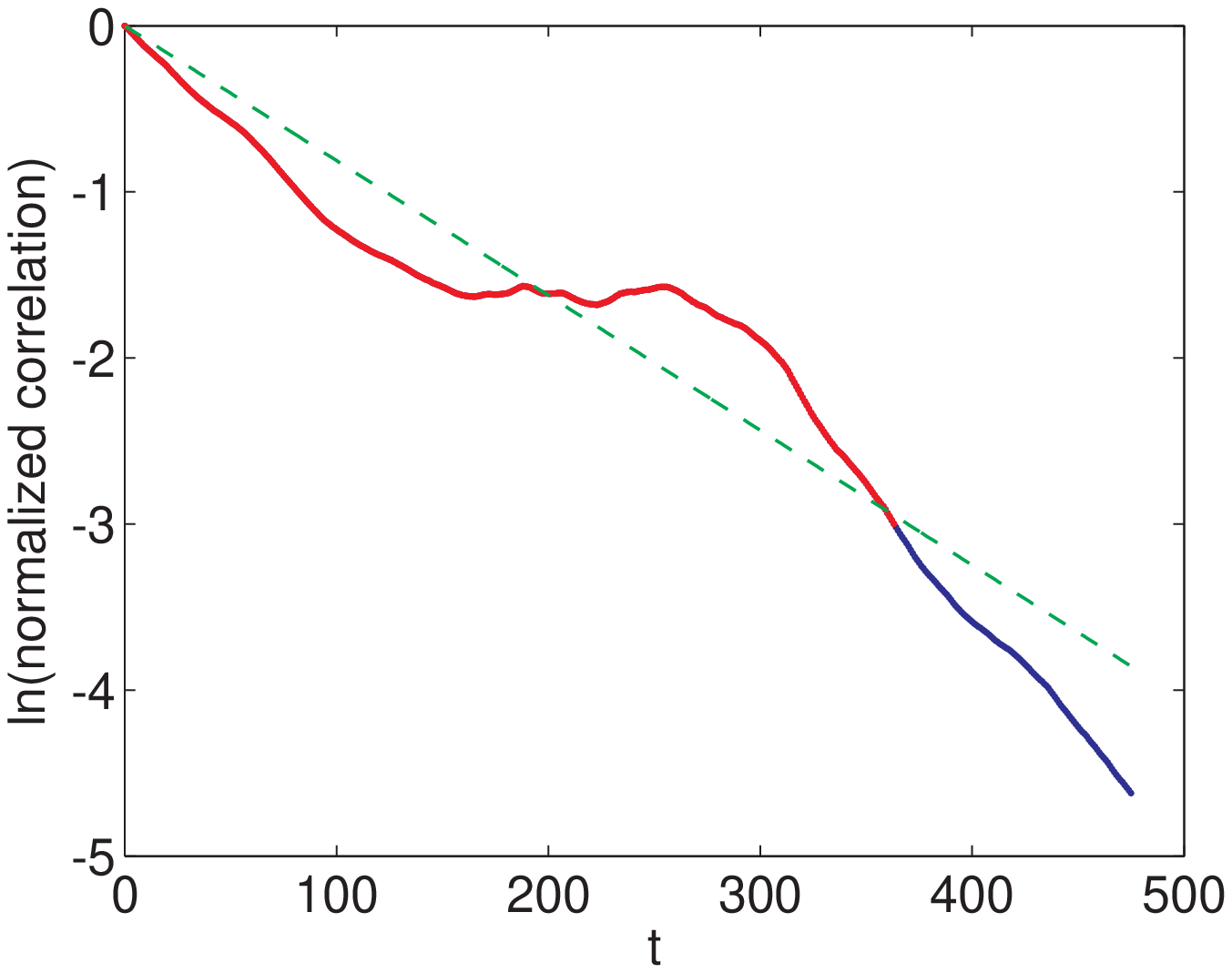} &
\includegraphics[width = \myFigureWidth \textwidth]{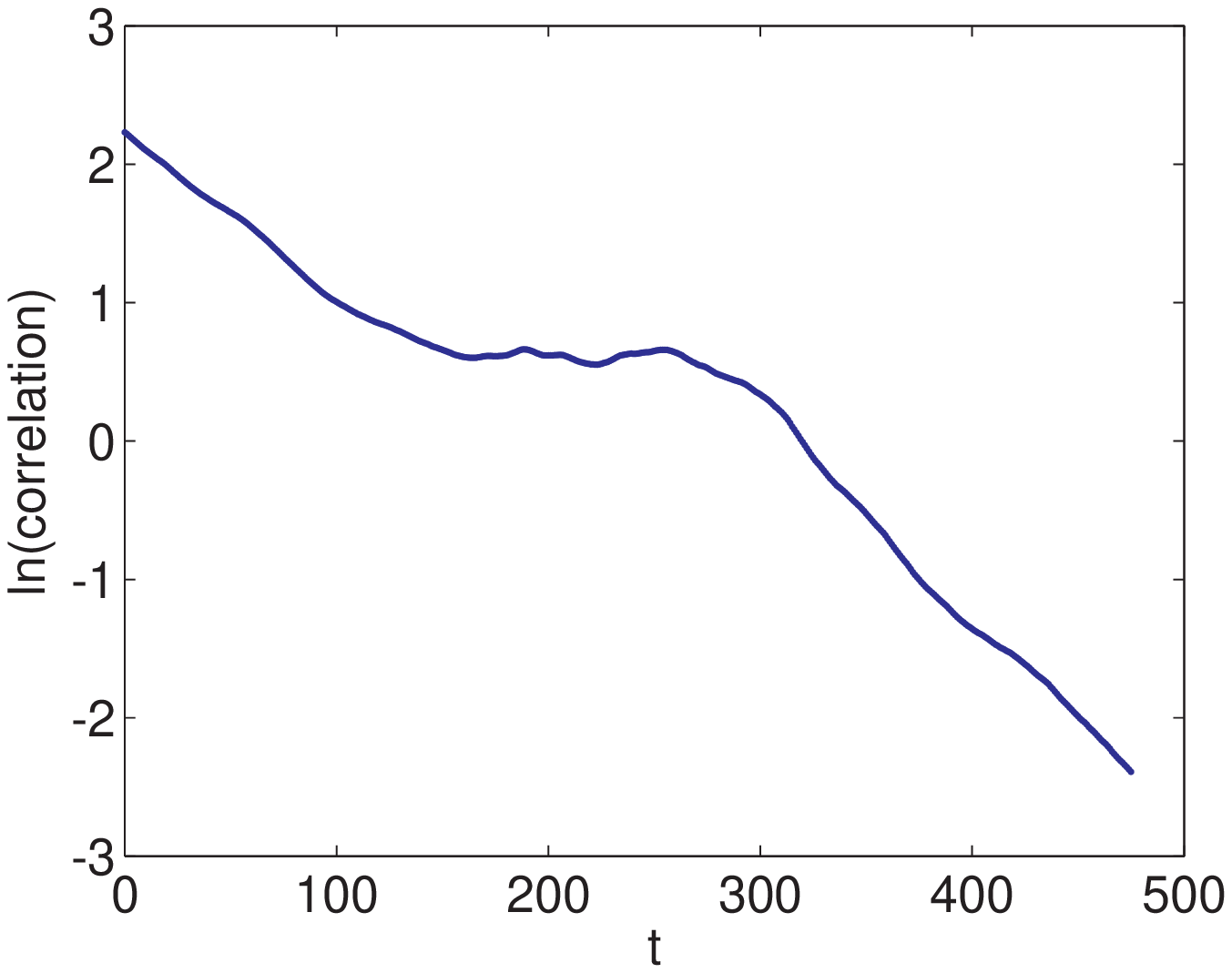} \\
\includegraphics[width = \myFigureWidth \textwidth]{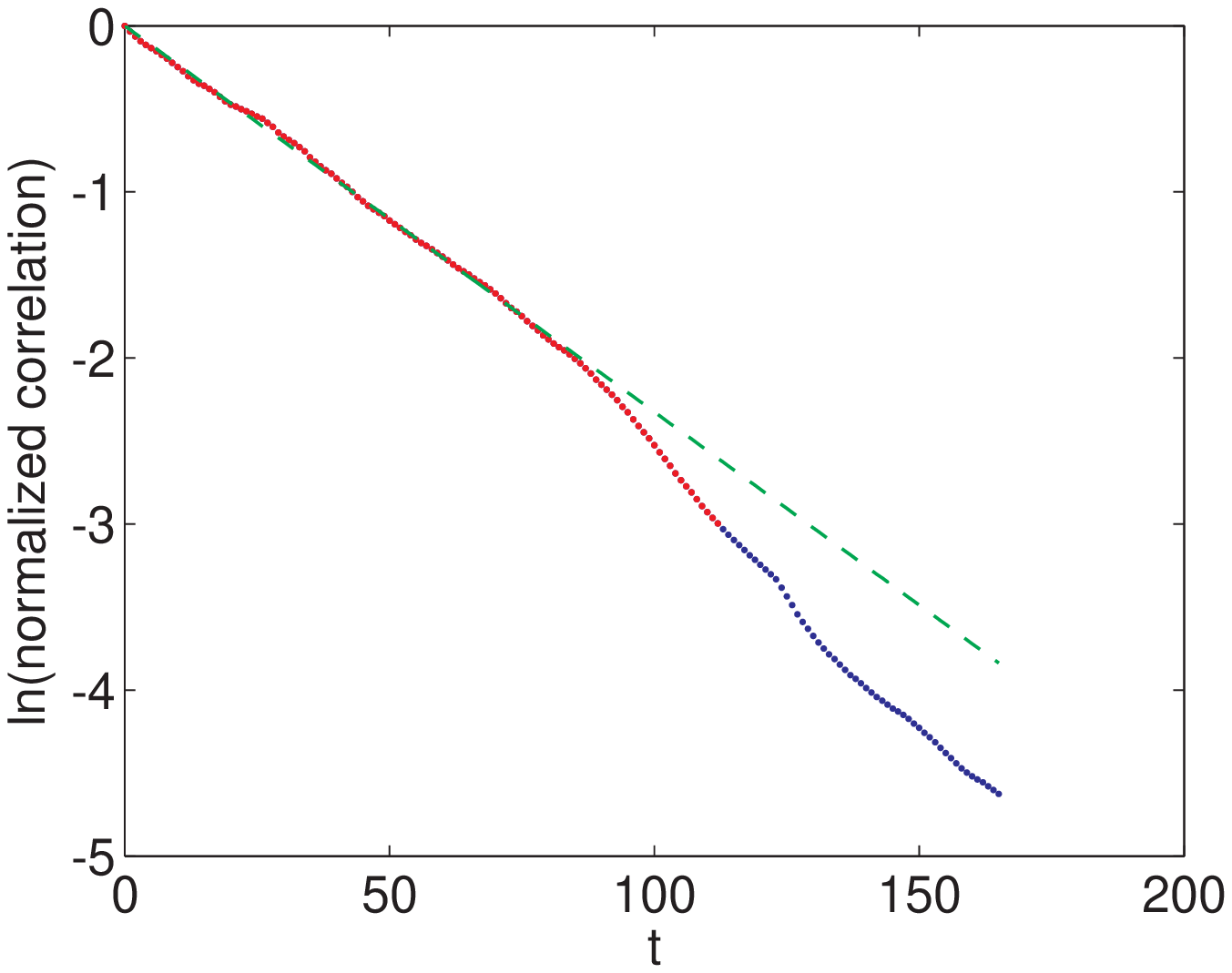} &
\includegraphics[width = \myFigureWidth \textwidth]{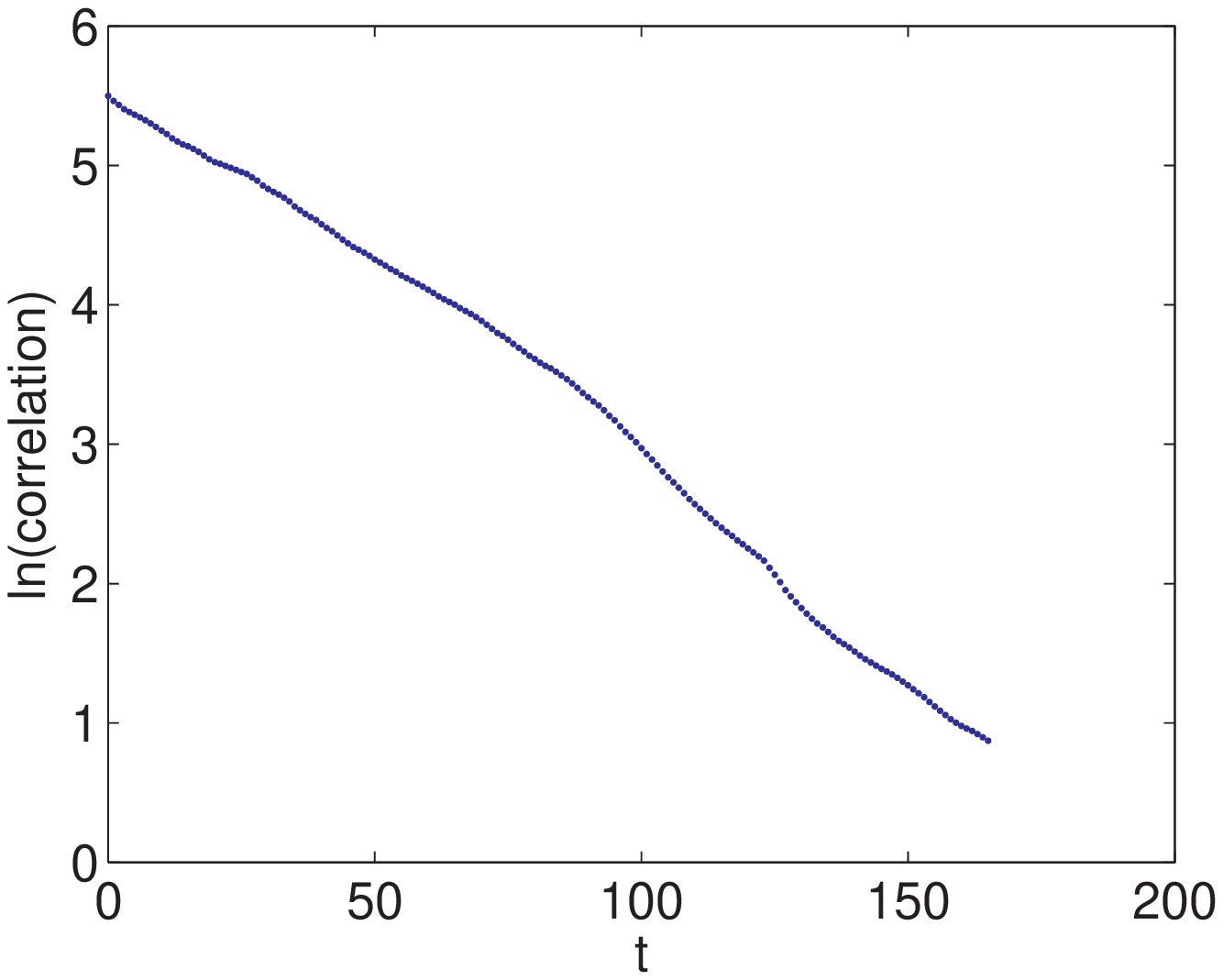} \\
\includegraphics[width = \myFigureWidth \textwidth]{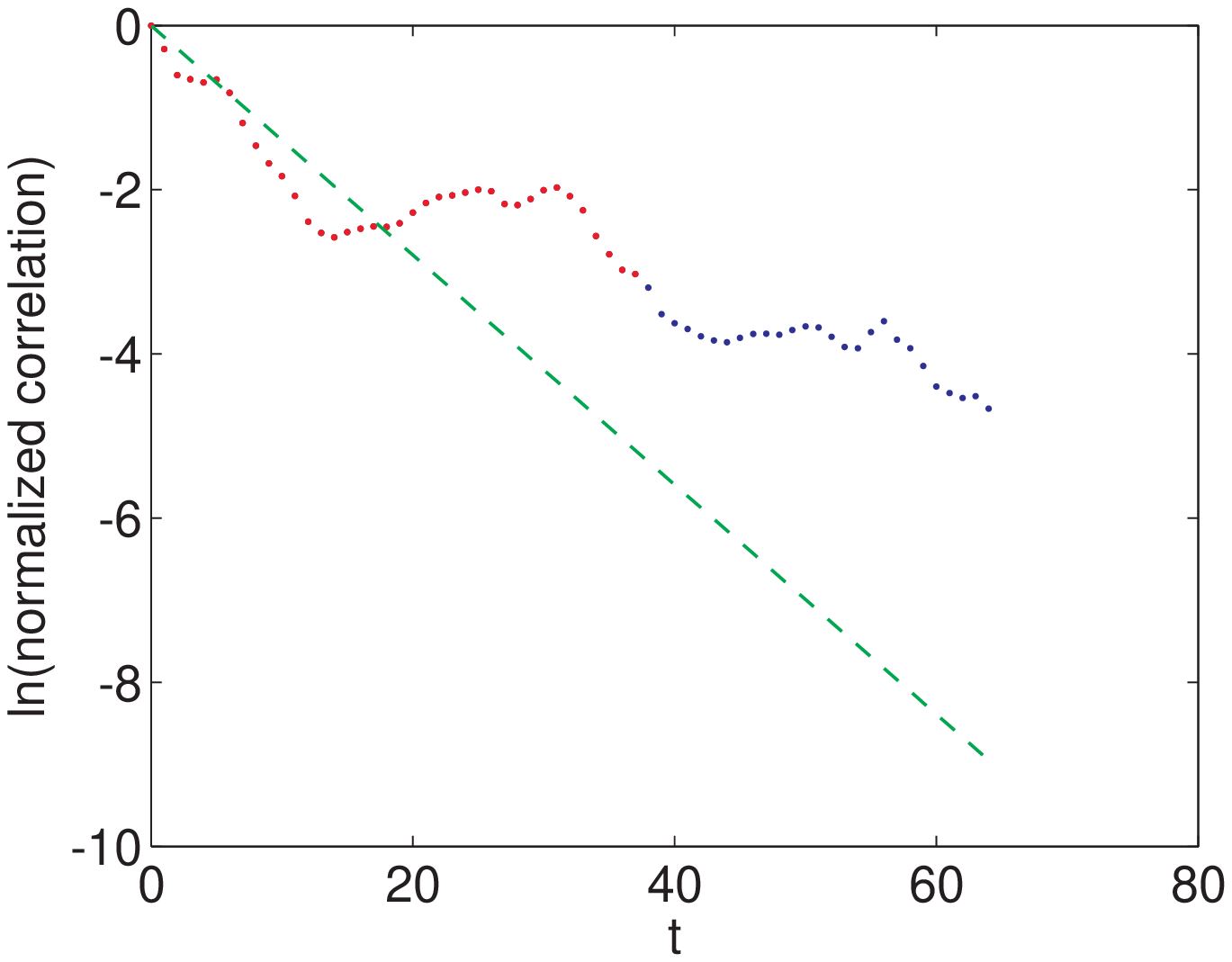} &
\includegraphics[width = \myFigureWidth \textwidth]{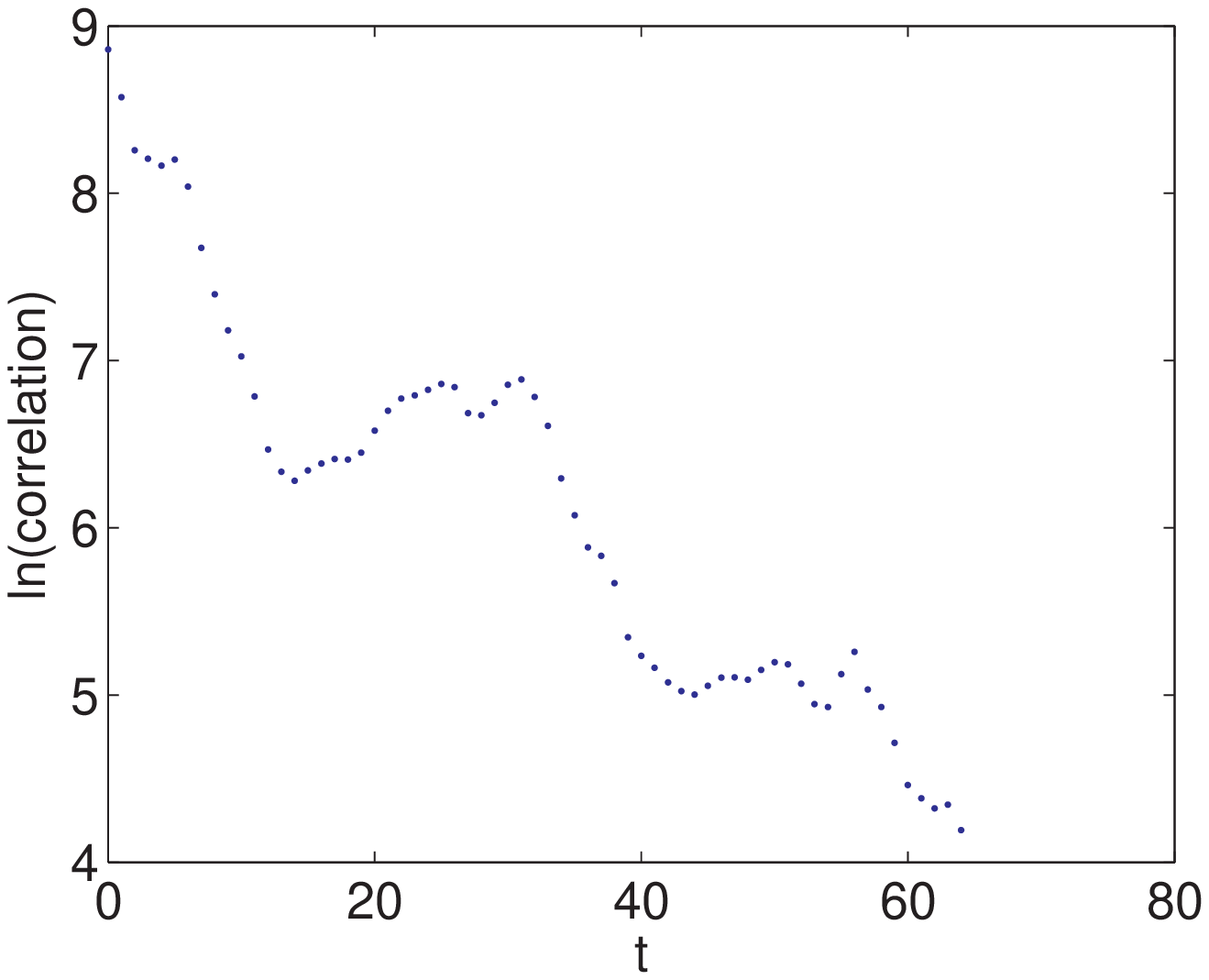} \\
\end{tabular}
\caption{Correlation functions computed from (\ref{correlationPaperItoProcess}) for $1.1\times 10^7$ time "ticks", with $10^3$ time steps between ticks, for $J=10^{-2}$ and, from top to bottom, $\sigma^2 = 1.09\times10^{-4}$, $1.02\times10^{-3}$, $0.96\times10^{-2}$, $1.04\times10^{-2}$, $1.14\times10^{-2}$, $3.47\times10^{-2}$ and $1.50\times10^{-1}$ respectively.}
\label{correlationCoefficient}
\end{figure}

The log plots of correlation function as a function of time $\tau$ are shown in Fig. \ref{correlationCoefficient}, with
the normalized one being fitted by a straight line whose slope, for the values shown,
is very close to $-J$. The third and fourth plots from the top are for $\sigma^2$ just below and just above $J$. It should
be noted, however, that for $J<\sigma^2$, the linearity of the plots generally deteriorate
rather dramatically for larger $\sigma^2$, as expected and observed in Fig. \ref{correlationCoefficient}.

\section{Cumulant Relaxation for Stochastic IG\lowercase{a} Process\label{CumulantRelaxationForStochasticIGaProcess}}

The number and dependence on parameters $J$ and $\sigma^2$ of time scales characterizing
relaxation is a function of the initial conditions. Here, we will concentrate on the longest
relaxation times. For the mean, it trivially follows from (\ref{correlationPaperItoProcess}) that 
\begin{equation}
\langle x\rangle = 1 + Ae^{-Jt}
\end{equation}
where $A$ is determined by the initial conditions: for instance, when all $x(0) = 0$,
$\langle x\rangle = 1 - e^{-Jt}$ and when all $x(0)=1$, $\langle x\rangle = 1$. Clearly, the relaxation process
will first establish the mean $\langle x\rangle = 1$ over the shortest time in the system, $J^{-1}$, which,
incidentally, does not depend on stochasticity. Consequently, in what follows, we
will omit the relaxation of the mean by using all $x(0)=1$ as the initial condition.

Using $\langle x\rangle = 1$ and, per (\ref{correlationPaperItoProcess}), 
\begin{eqnarray}
% \begin{equation}
\mathrm{d}(x^2) 
&& = 2x\mathrm{d}x+\mathrm({d}x)^2 
\nonumber \\ && 
= 2x\left[ -J(x-1)\mathrm{d}t + \sqrt{2}\sigma x\mathrm{d}B \right] + 2\sigma^2 x^2 \mathrm{d}t
% \end{equation}
\end{eqnarray}
we find for the variance (second cumulant) $\kappa_2 = \langle x^2\rangle - \langle x\rangle^2 = \langle x^2 \rangle - 1$ 
\begin{equation}
\mathrm{d}\kappa_2 = \mathrm{d} \langle x^2 \rangle = 2\left[ (-J+\sigma^2)\kappa_2 + \sigma^2 \right] \mathrm{d}t
\label{correlationPaperDXSquare}
\end{equation}
and, under assumption $J>\sigma^2$, 
\begin{equation}
\label{correlationPaperKappa2}
\kappa_2 = \frac{1}{\frac{J}{\sigma^2} - 1} + D \cdot e^{-2(J-\sigma^2)t}
\end{equation}
where $D$ is determined by the initial conditions: for instance, when all $x(0)=1$, that
is $\kappa_2(0) = 0$ 
\begin{equation}
\label{correlationPaperKappa2AtZero}
\kappa_2 = \frac{1 - e^{-2(J-\sigma^2)t}}{\frac{J}{\sigma^2} - 1}
\end{equation}
and when $\kappa_2(0)= \sigma^2/(J-\sigma^2)$, that is the variance is already that of the steady-state distribution, we find $D=0$, as it ought. Clearly, (\ref{correlationPaperCorrelationFunctionInSteadyState}) and (\ref{correlationPaperKappa2AtZero}) are the limiting cases of (\ref{correlationPaperIGaCorrelationFunction}) for $t\rightarrow \infty$ and $\tau = 0$ respectively. From (\ref{correlationPaperKappa2AtZero}) it is obvious that as $\sigma^2 \xrightarrow{>}J$, the relaxation time diverges as $(J-\sigma^2)^{-1}$, as does $\kappa_2$. 

Similarly, as $2\sigma^2 \xrightarrow{>}J$ the relaxation time diverges as $(J-2\sigma^2)^{-1}$, as does $\kappa_3$, the
third cumulant: 
\begin{eqnarray}
% \begin{widetext}
% \begin{equation}
% \begin{split}
\kappa_3 
% & 
& = & \frac{1}{\left( \frac{J}{\sigma^2}-1 \right) \left( \frac{J}{\sigma^2}-2 \right) \left( \frac{J}{\sigma^2}-4 \right) } \cdot
\nonumber \\ && 
\left[ 4\left( \frac{J}{\sigma^2}-4 \right) - 12\left( \frac{J}{\sigma^2}-2 \right) e^{-2(J-\sigma^2)t} \right. 
\nonumber \\ && 
\quad \left. + 8\left( \frac{J}{\sigma^2}-1 \right) e^{-3(J-2\sigma^2)t} \right]
% \\
% & 
\nonumber \\ &
\approx & \frac{4\left( 1-e^{-3(J-2\sigma^2)t} \right)}{\frac{J}{\sigma^2} - 2}
\label{correlationPaperKappa3}
% \end{split}
% \end{equation}
% \end{widetext}
\end{eqnarray}
Both (\ref{correlationPaperKappa2AtZero}) and (\ref{correlationPaperKappa3}) are particular cases of (\ref{correlationPaperTimeDependenceCummulants}) and (\ref{correlationPaperTimeDependenceCummulantsLambda}).

\begin{figure*}[!htbp]
\centering
\begin{tabular}{ccc}
\includegraphics[width = \myFigureWidth \textwidth]{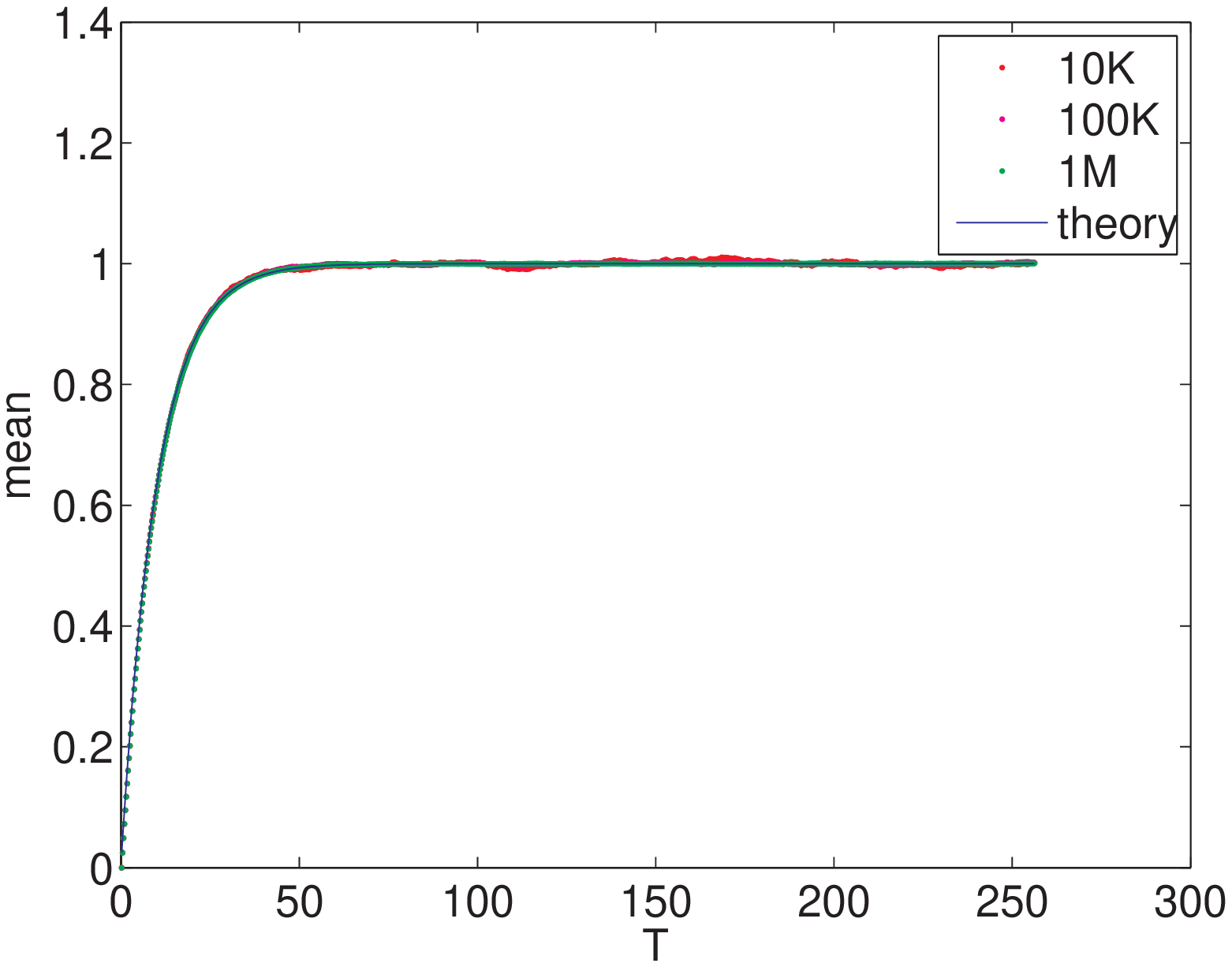} &
\includegraphics[width = \myFigureWidth \textwidth]{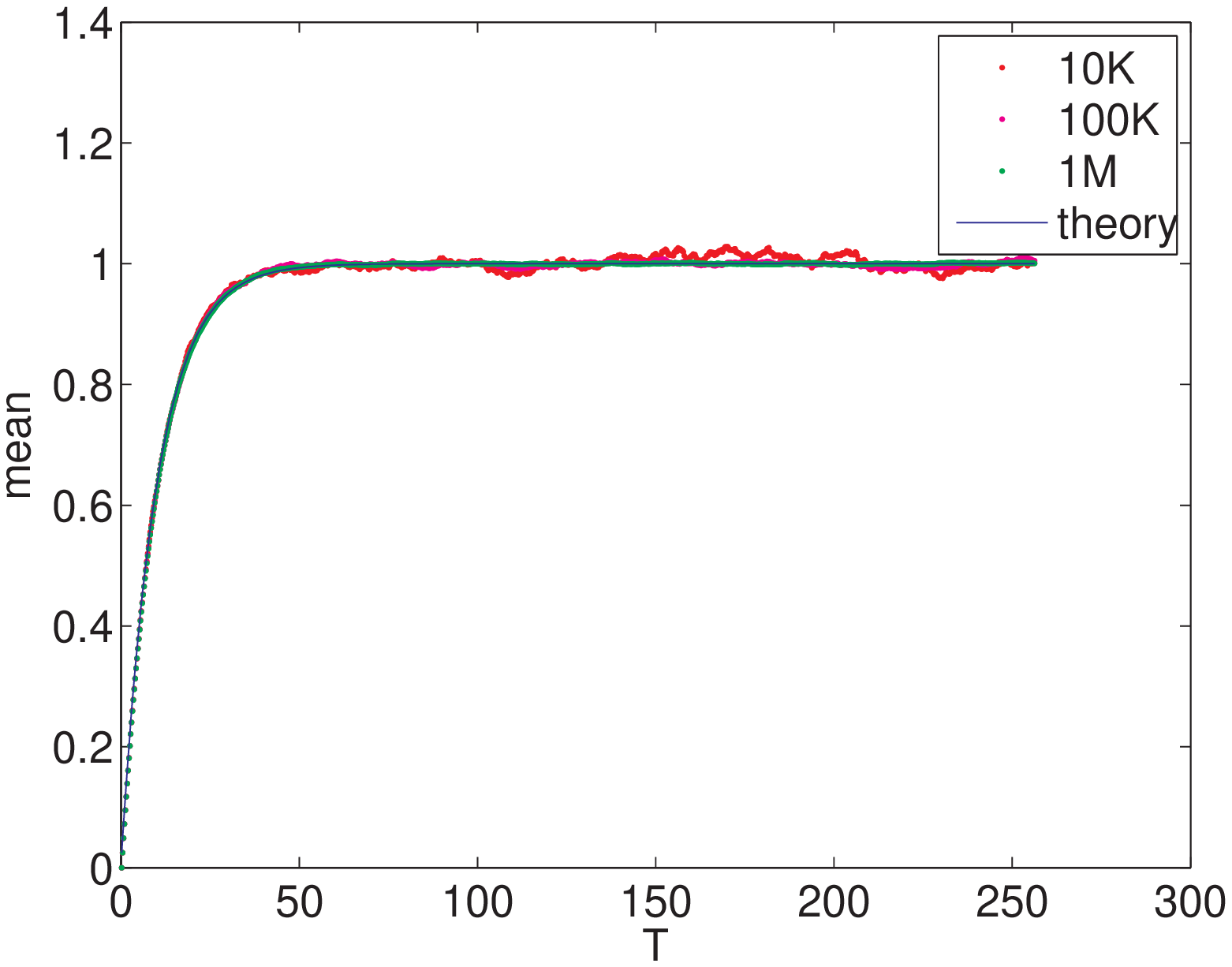} &
\includegraphics[width = \myFigureWidth \textwidth]{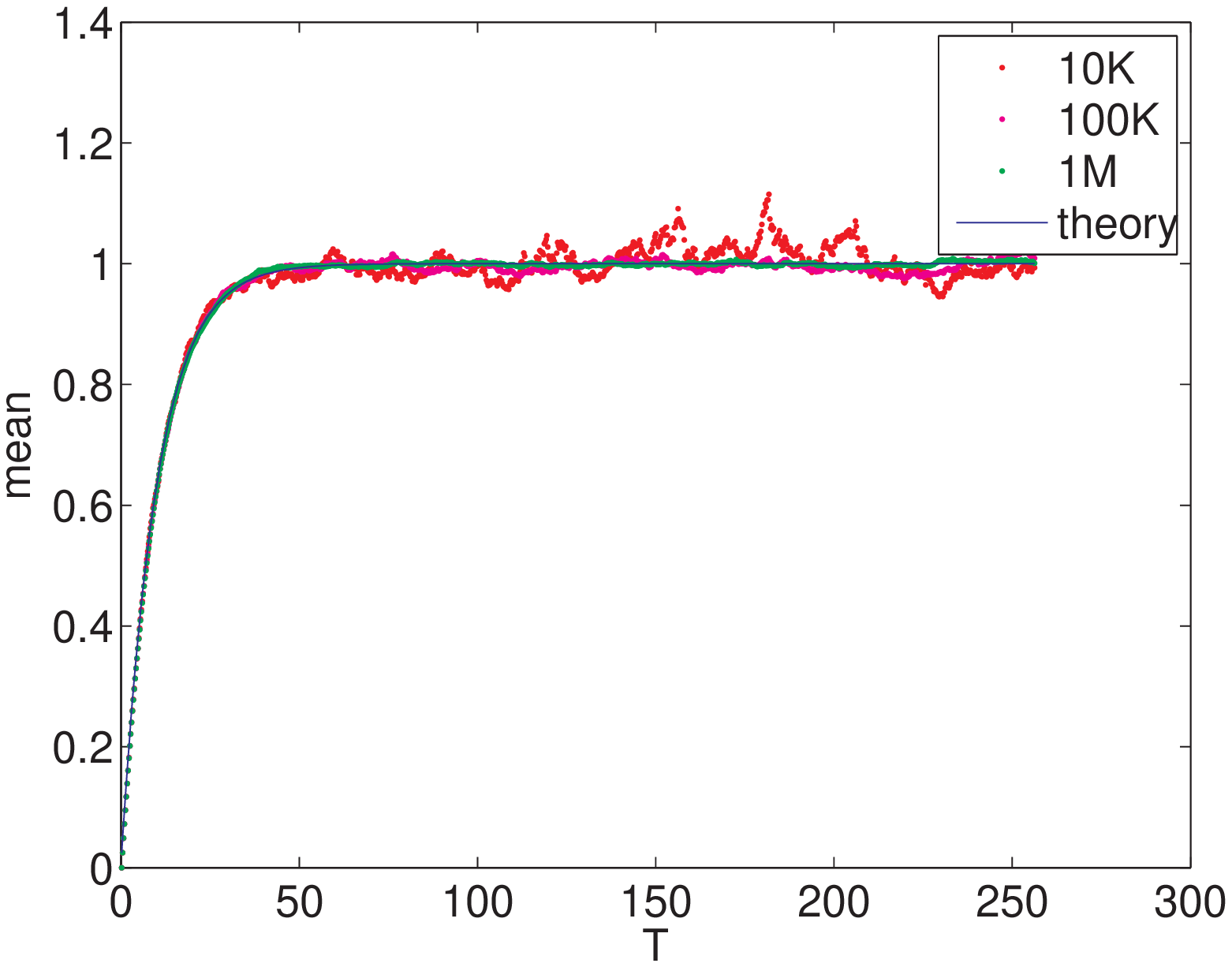} \\
\end{tabular}
\caption{Saturation of the mean vis-$\mathrm{\grave{a}}$-vis theoretical result, $\langle x\rangle = 1 - e^{-Jt}$, for $x(0)=0$ for $10^4$, $10^5$, and $10^6$ time series respectively, here $J=10^{-1}$ and, from left to right, $\sigma^2=10^{-2}$, $5\times10^{-2}$, and $1.1\times10^{-1}$ respectively.}
\label{meanTrend}
\end{figure*}

\begin{figure*}[!htbp]
\centering
\begin{tabular}{ccc}
\includegraphics[width = \myFigureWidth \textwidth]{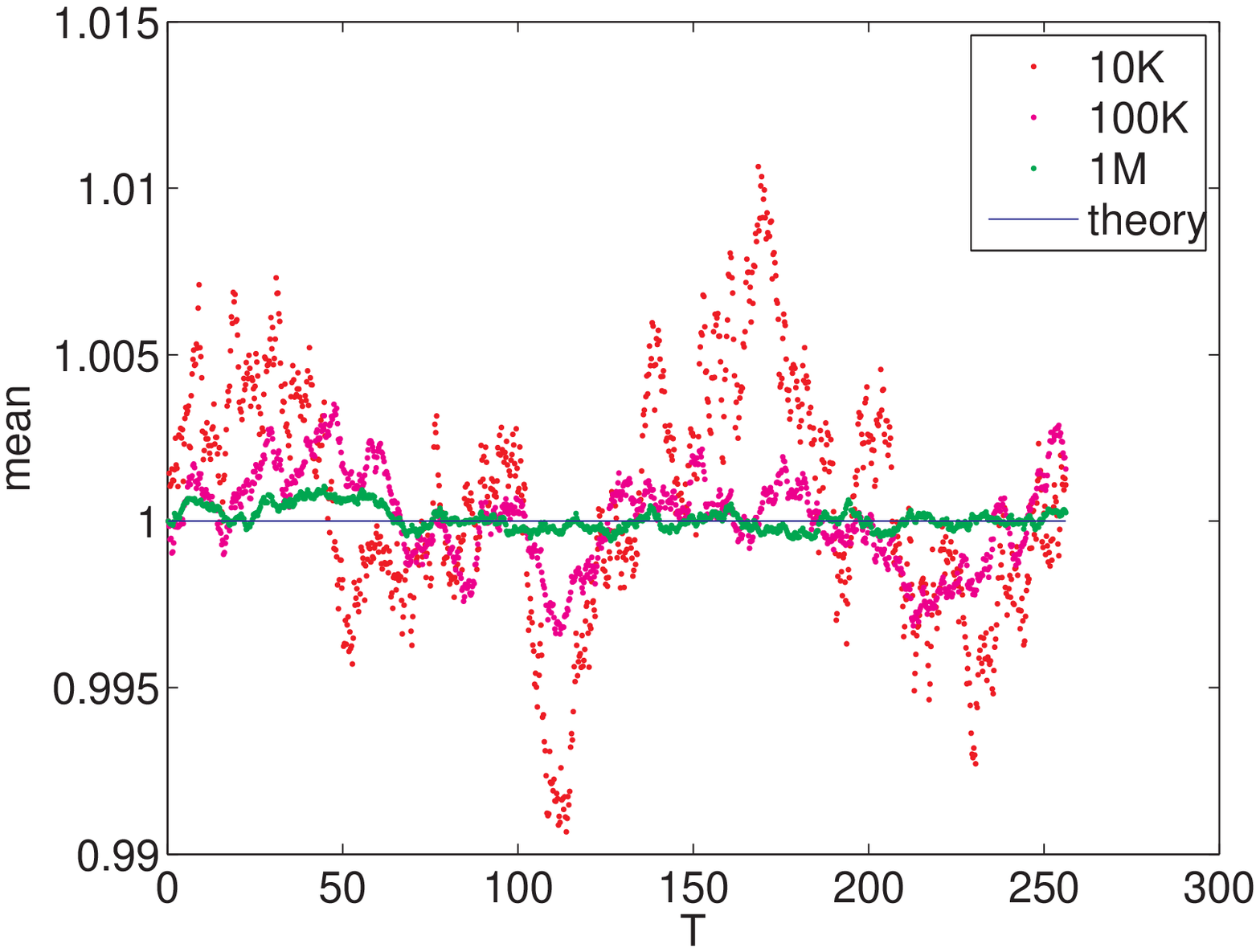} &
\includegraphics[width = \myFigureWidth \textwidth]{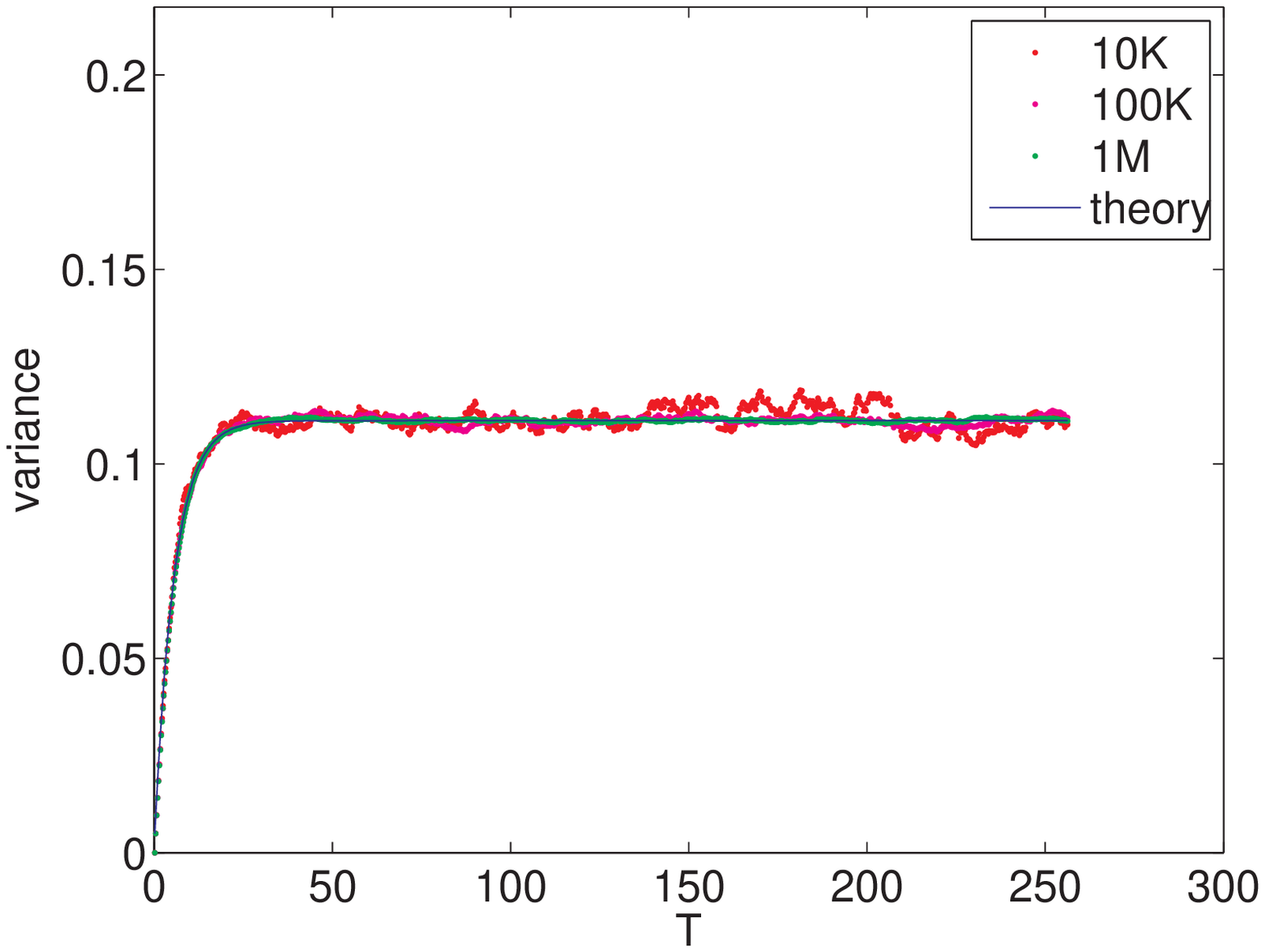} &
\includegraphics[width = \myFigureWidth \textwidth]{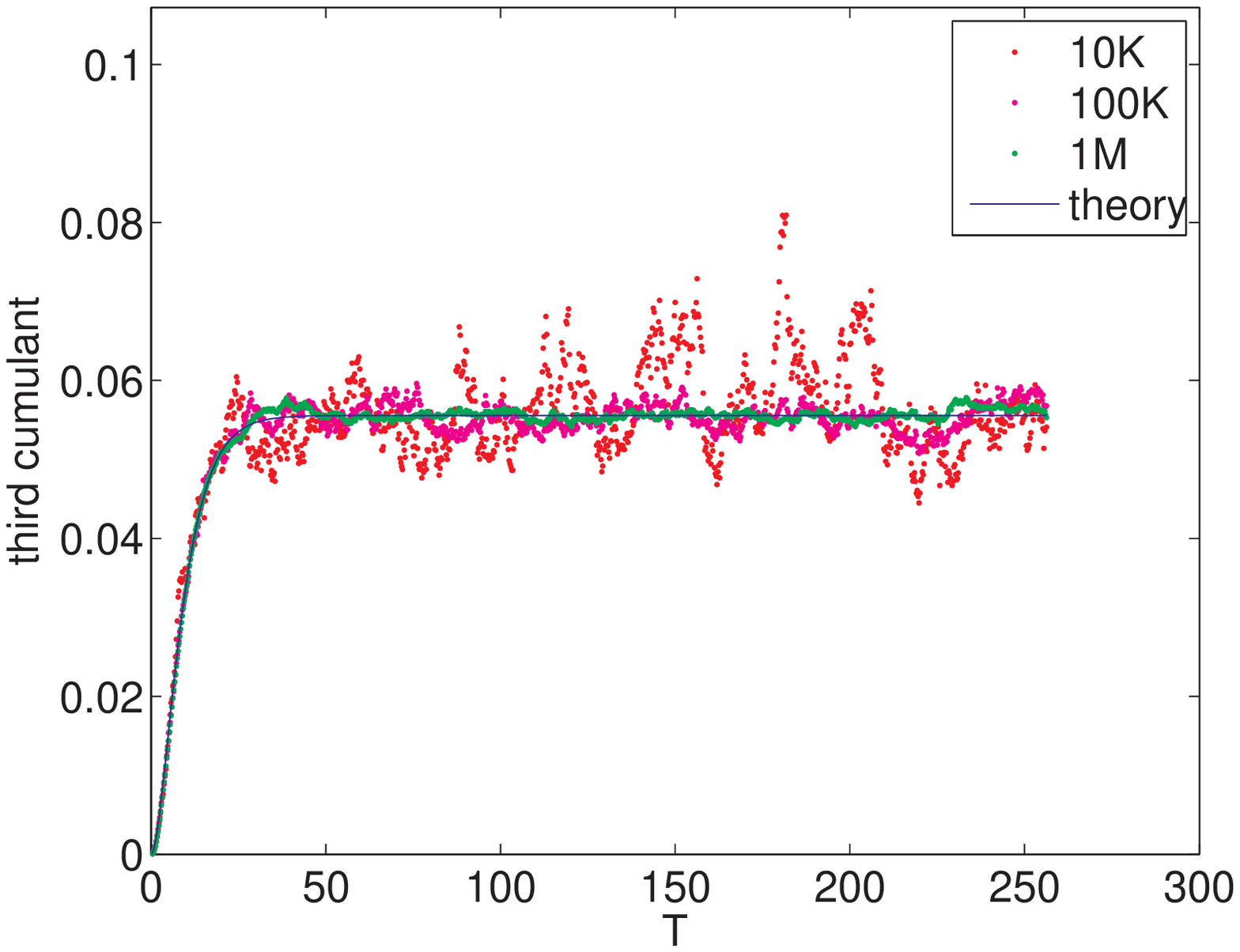} \\
\includegraphics[width = \myFigureWidth \textwidth]{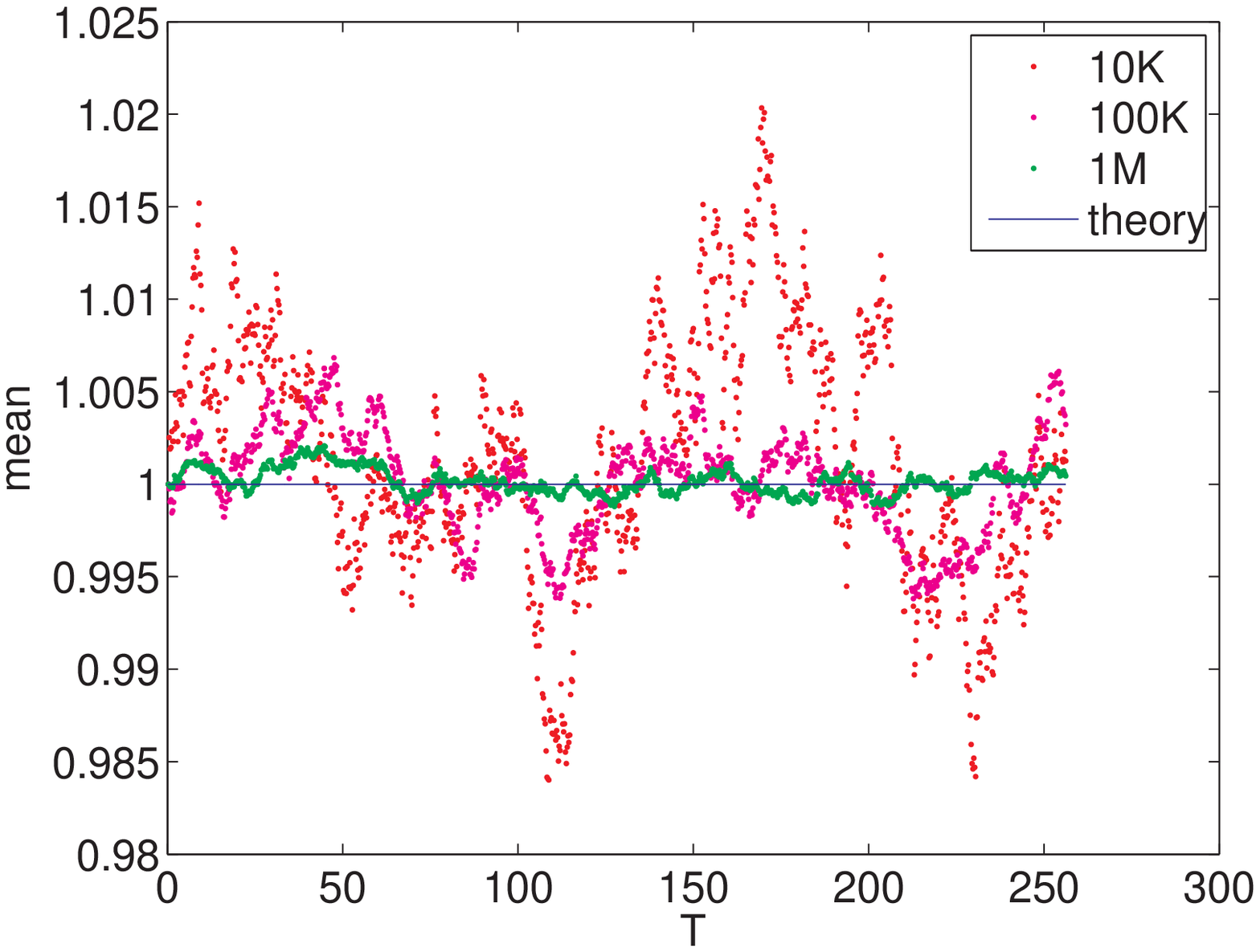} &
\includegraphics[width = \myFigureWidth \textwidth]{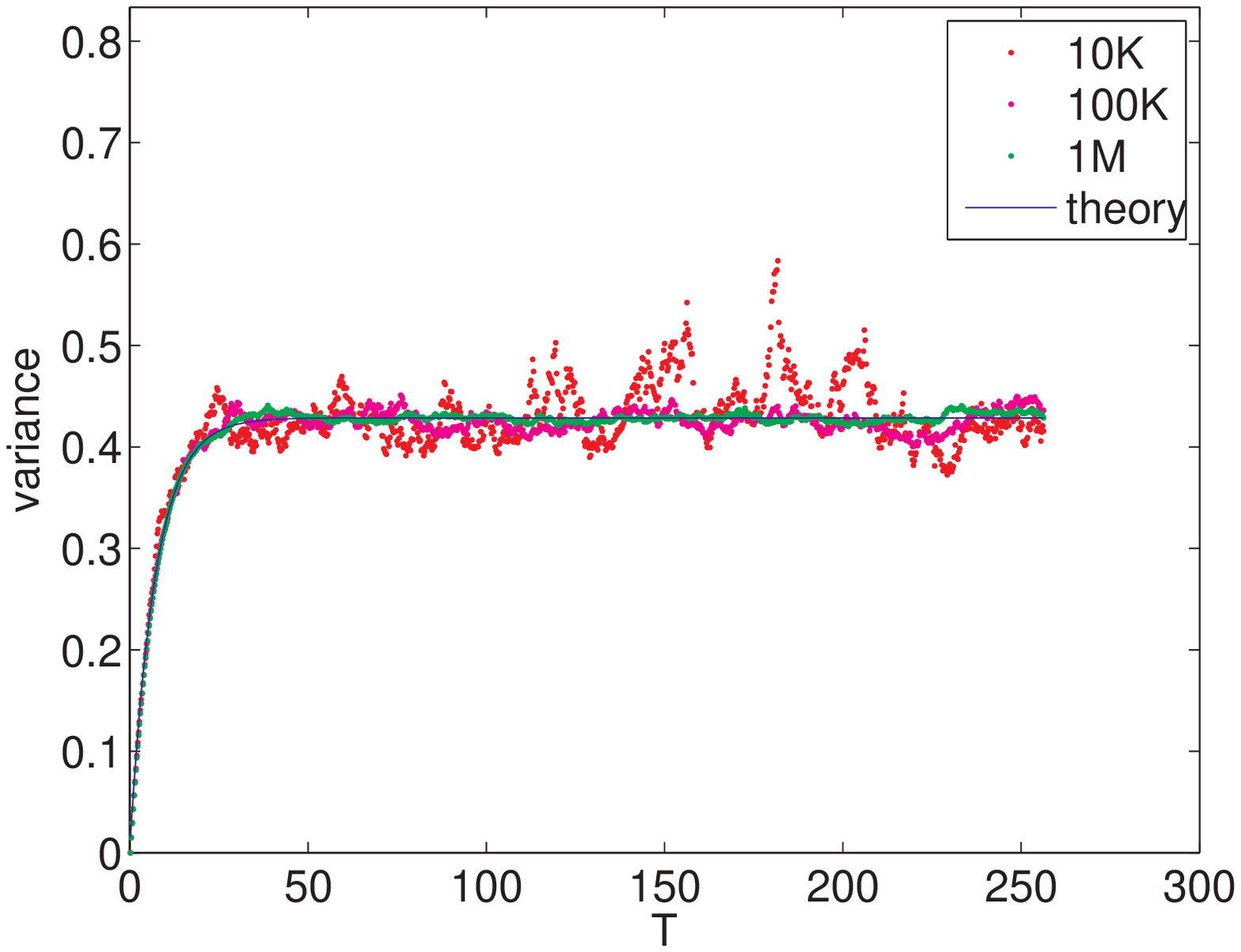} &
\includegraphics[width = \myFigureWidth \textwidth]{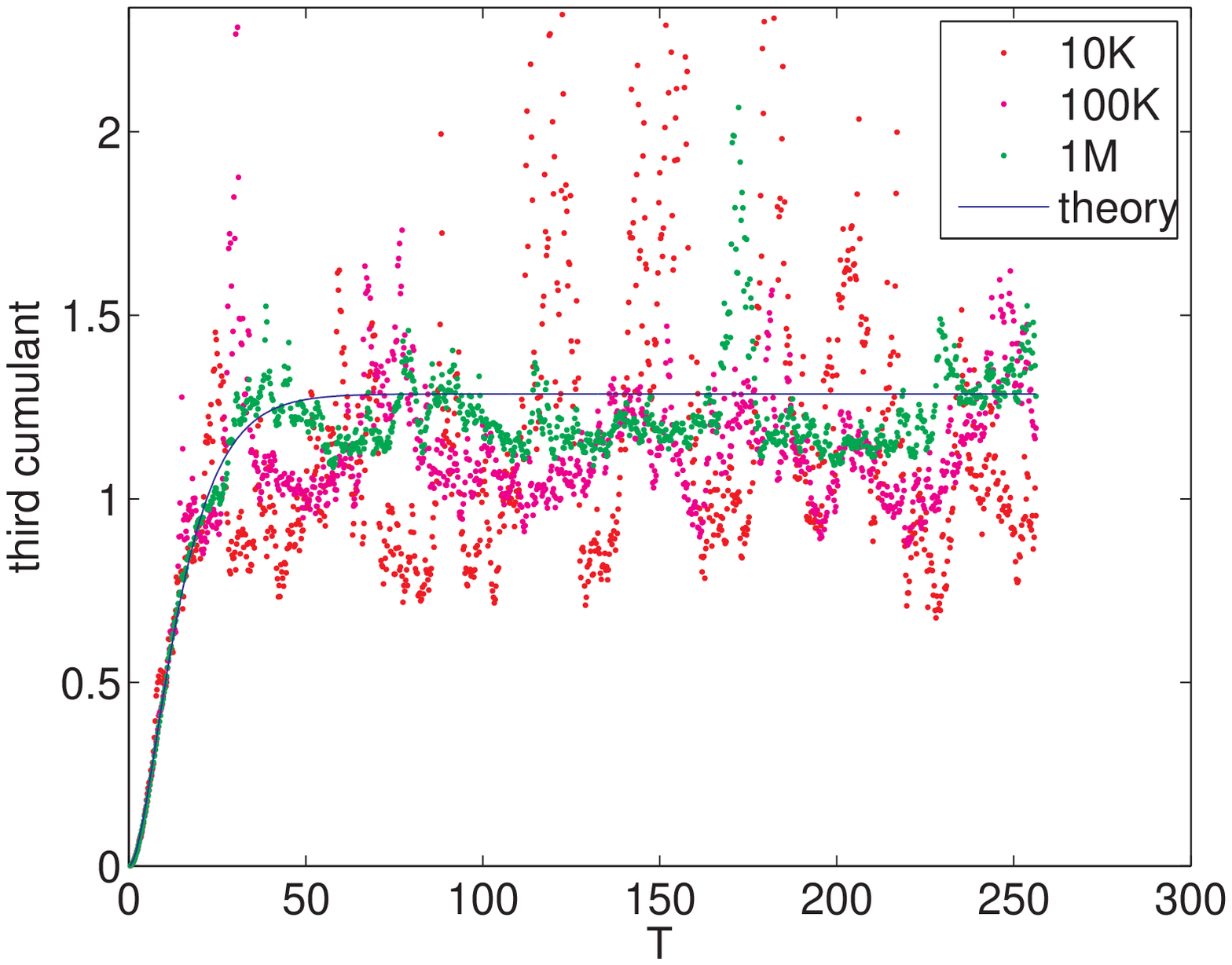} \\
\includegraphics[width = \myFigureWidth \textwidth]{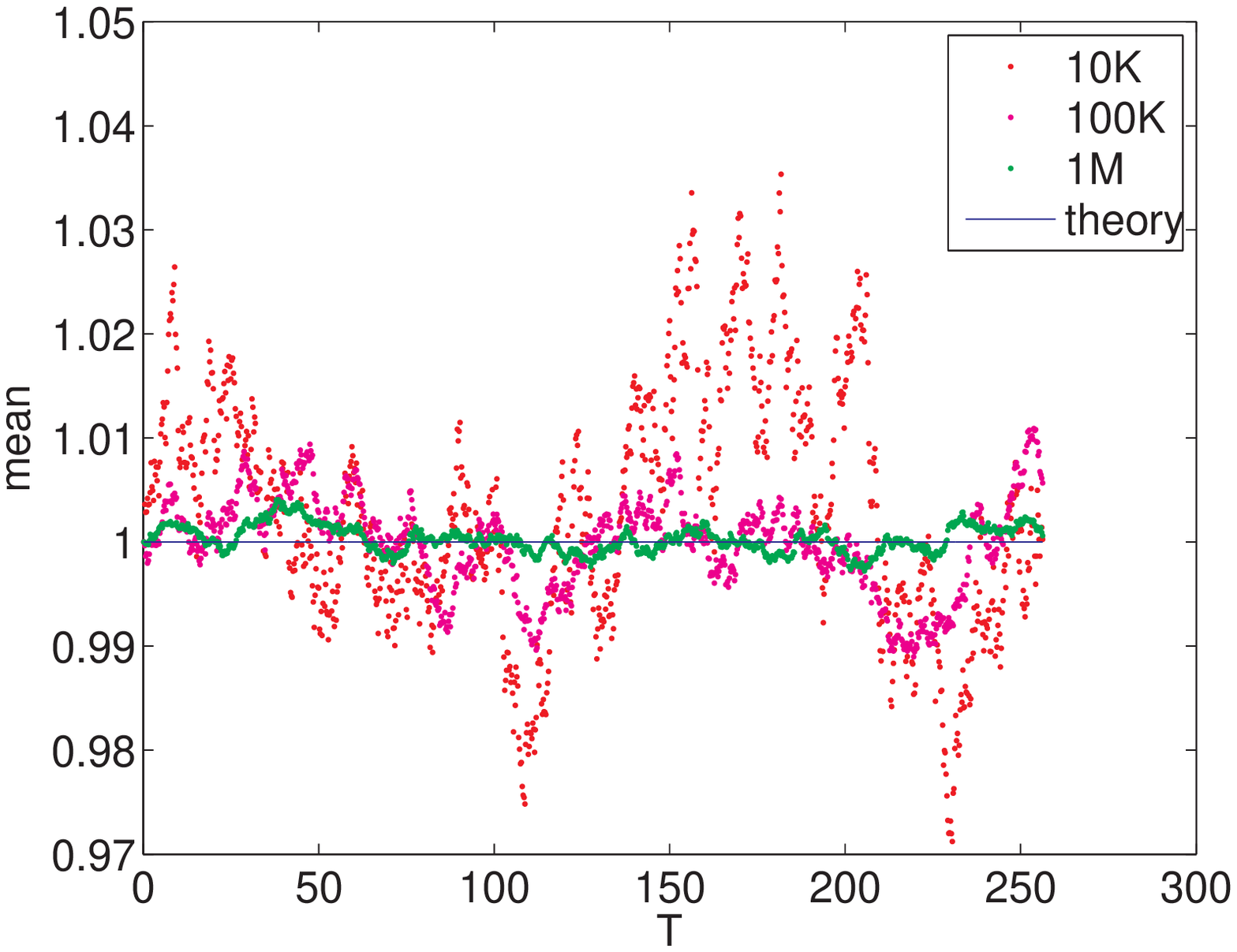} &
\includegraphics[width = \myFigureWidth \textwidth]{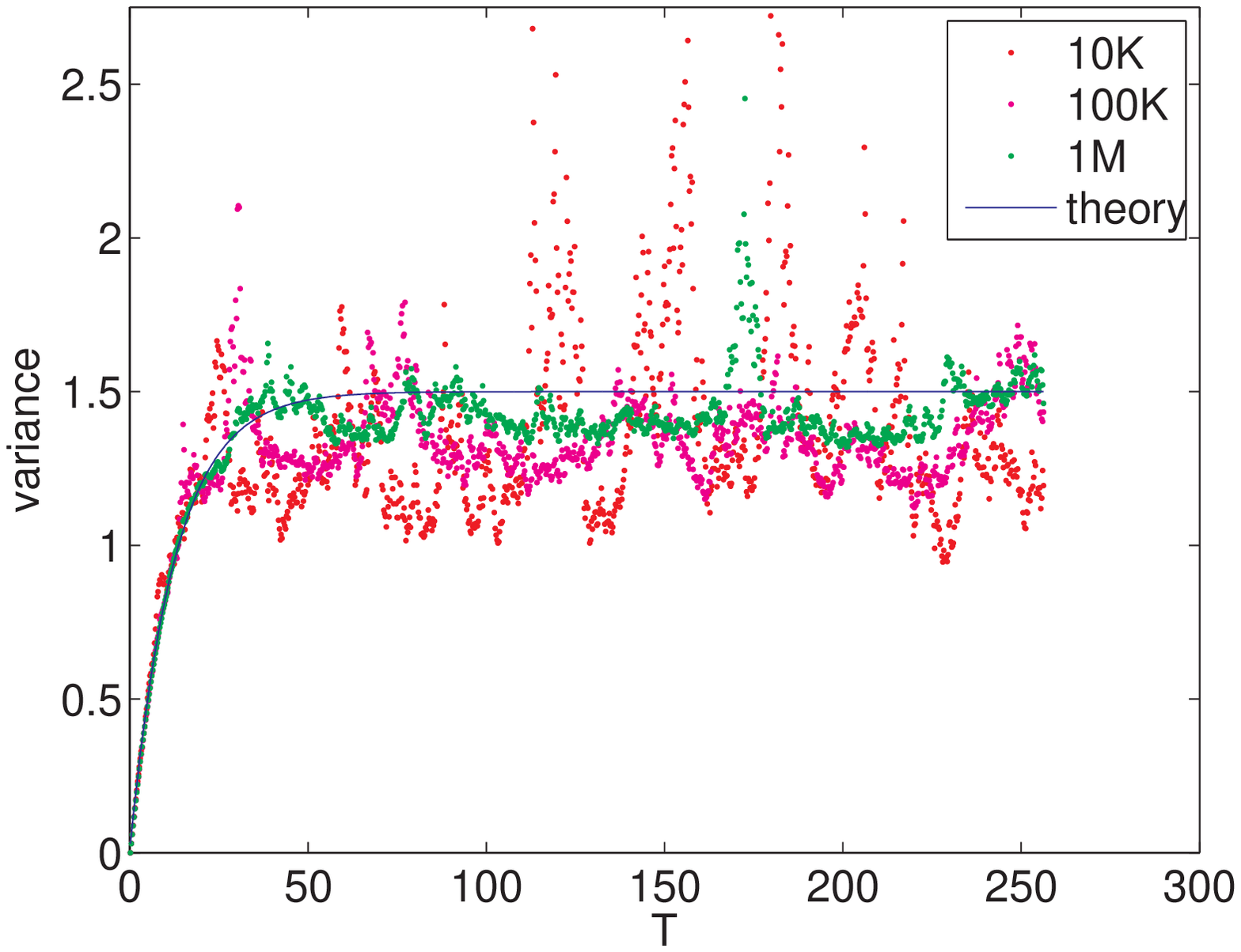} &
\includegraphics[width = \myFigureWidth \textwidth]{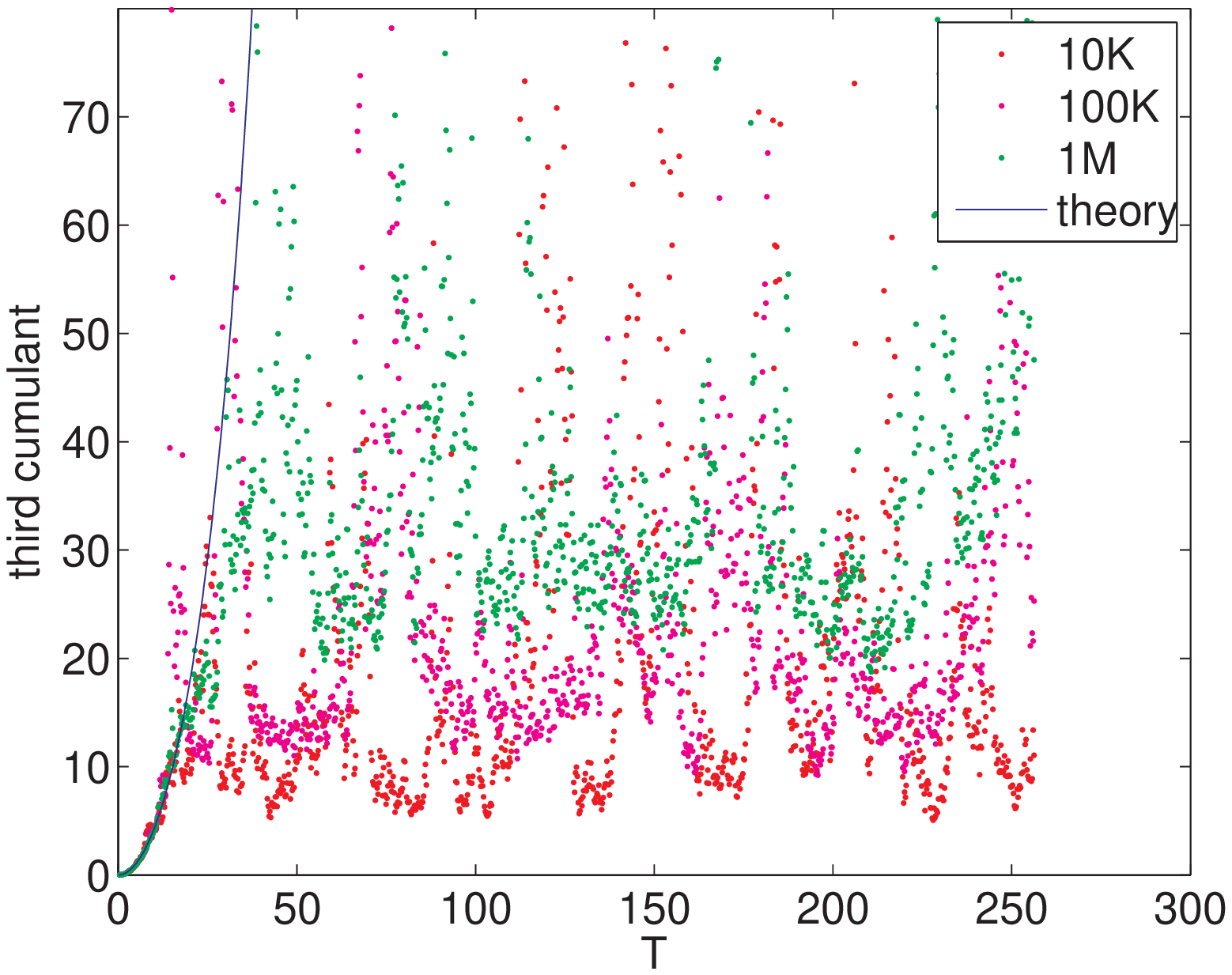} \\
\includegraphics[width = \myFigureWidth \textwidth]{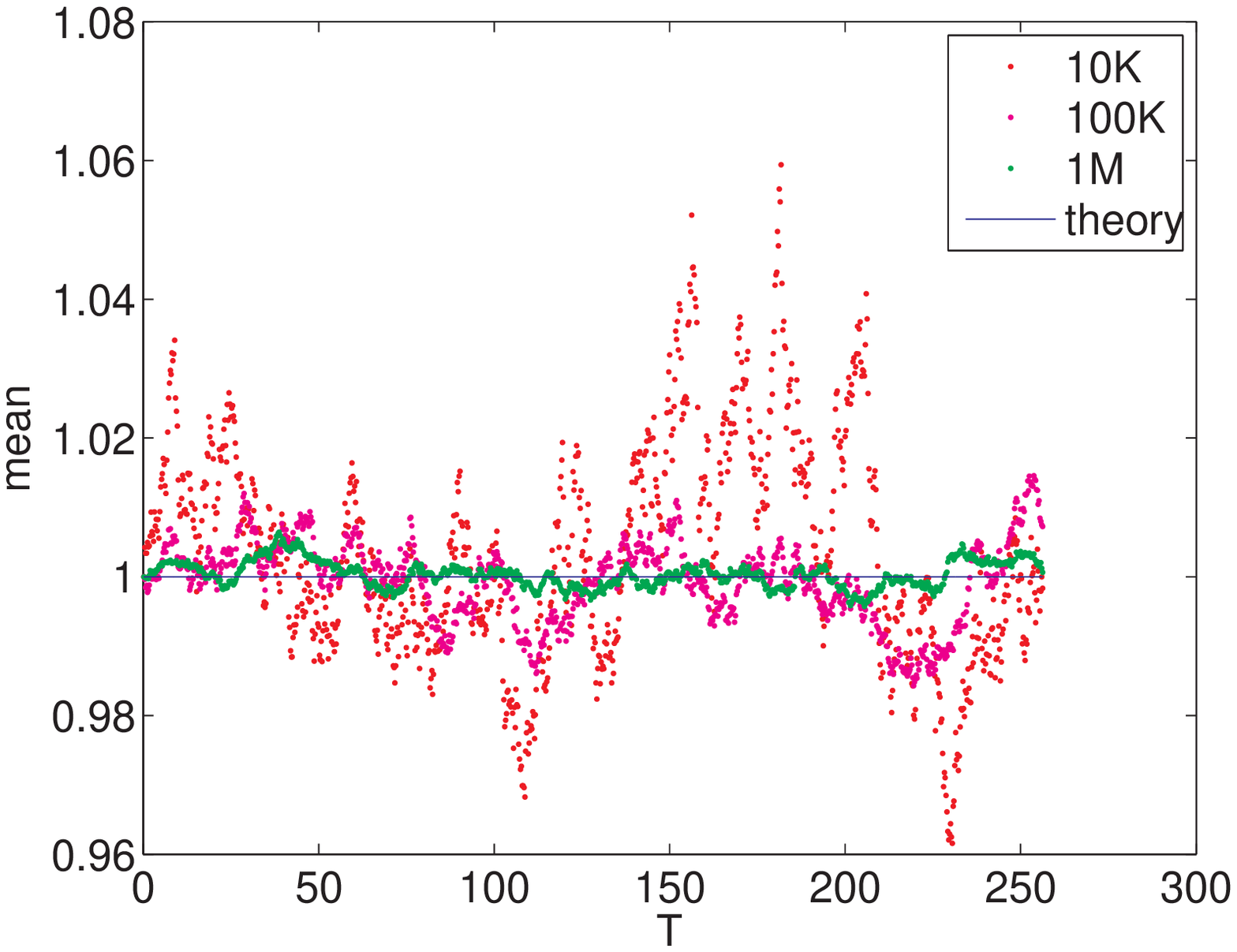} &
\includegraphics[width = \myFigureWidth \textwidth]{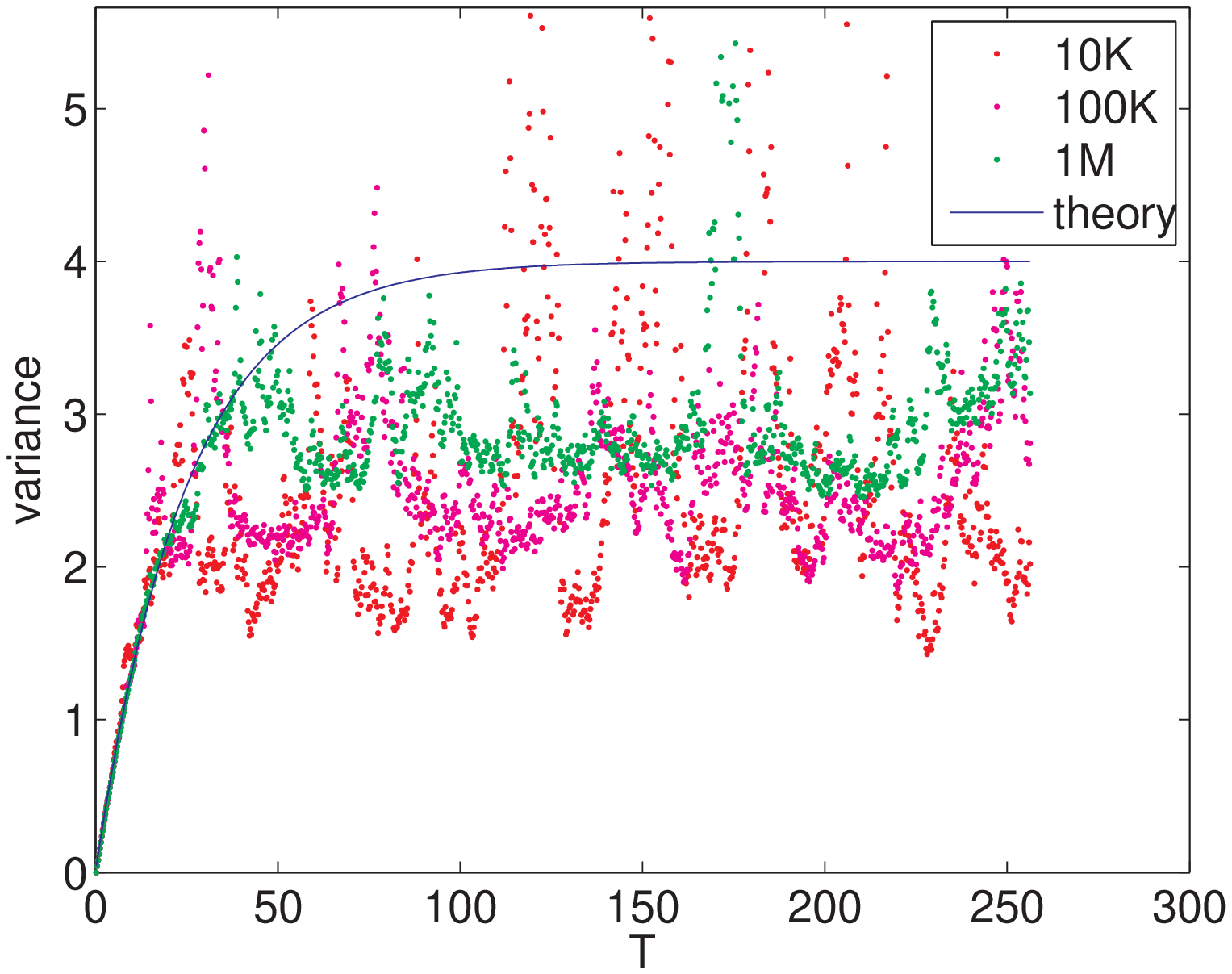} &
\includegraphics[width = \myFigureWidth \textwidth]{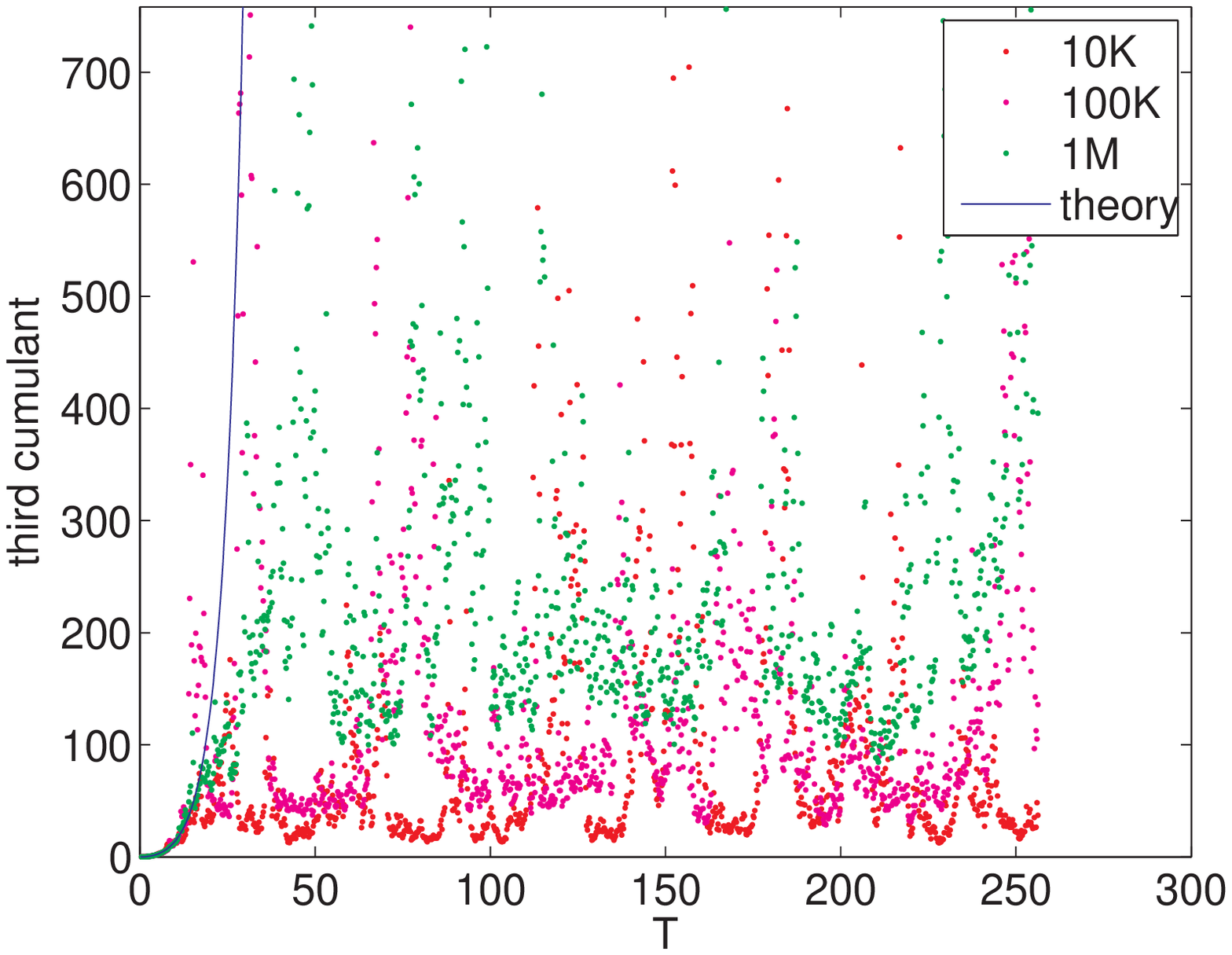} \\
\includegraphics[width = \myFigureWidth \textwidth]{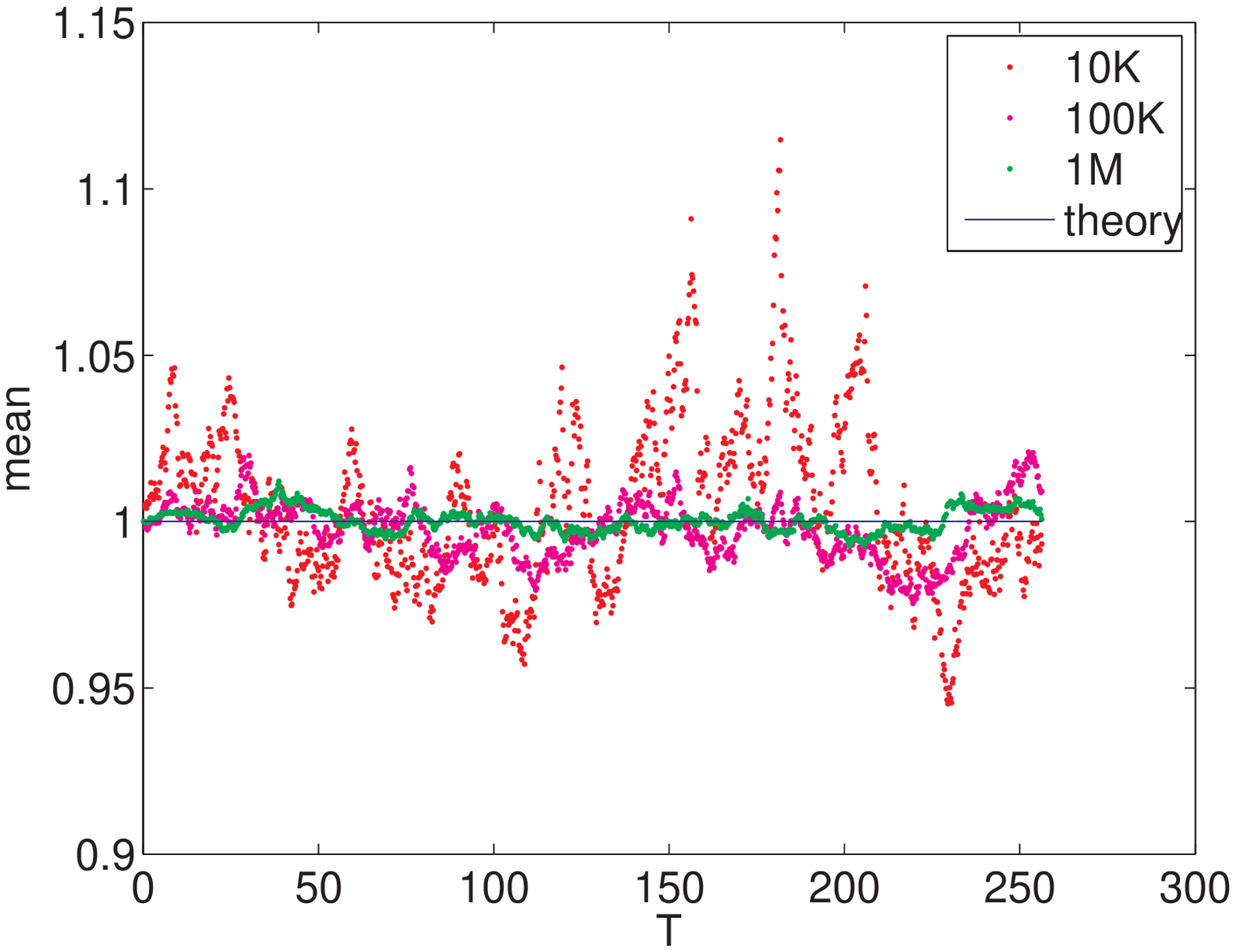} &
\includegraphics[width = \myFigureWidth \textwidth]{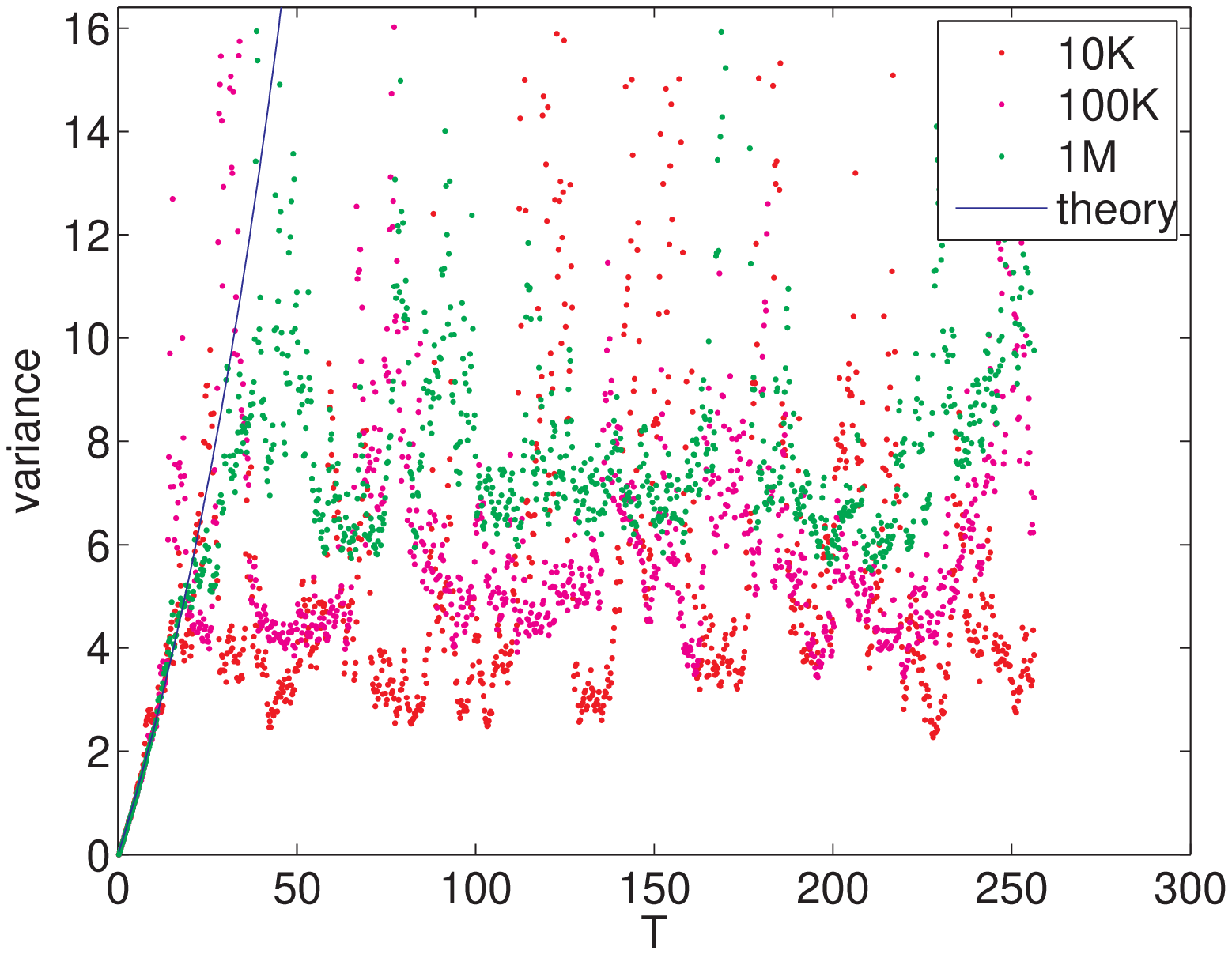} &
\includegraphics[width = \myFigureWidth \textwidth]{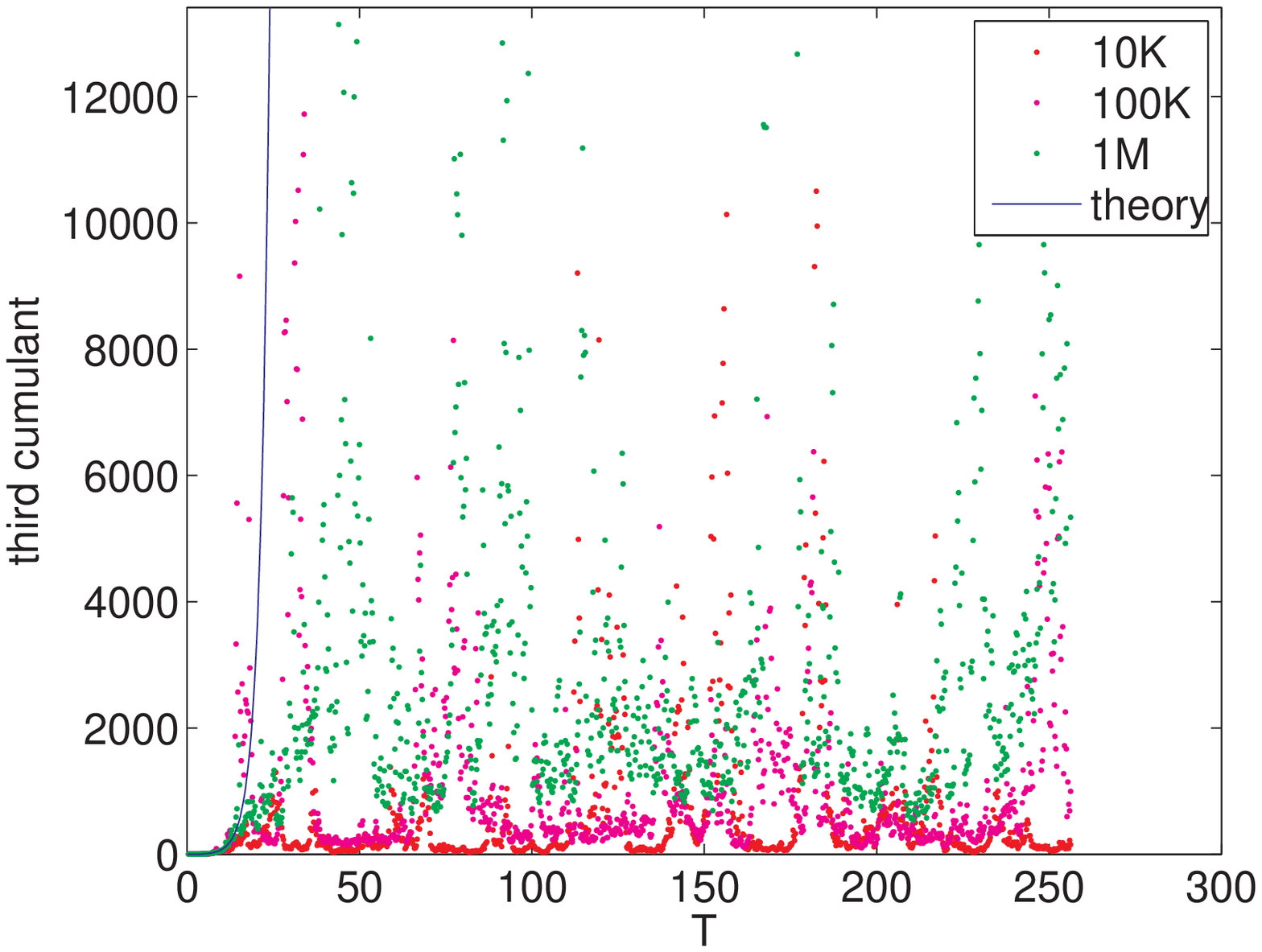} \\
\end{tabular}
\caption{Saturation of the mean, variance, and third cumulant vis-$\mathrm{\grave{a}}$-vis theoretical result, $\langle x\rangle = 1$, (\ref{correlationPaperKappa2AtZero}) and (\ref{correlationPaperKappa3}) for $x(0)=1$ for $10^4$, $10^5$, and $10^6$ time series respectively, here $J=10^{-1}$ and, from top to bottom, $\sigma^2=10^{-2}$, $3\times10^{-2}$, $6\times10^{-2}$, $8\times10^{-2}$, and $1.1\times10^{-1}$ respectively.}
\label{meanVarianceSkewTrend}
\end{figure*}

To verify our results, we numerically generate a large number of time series (\ref{correlationPaperItoProcess})
($10^4$, $10^5$, and $10^6$ respectively) and evaluate the cumulants at each of the $256$ 
consecutive time "ticks"; there are $2^{10}$ time steps between the ticks. Except for a
single illustration for the mean, where we used $x(0)=0$, we use the same initial
condition $x(0)=1$. The relaxation results are shown in Figs. \ref{meanTrend} and \ref{meanVarianceSkewTrend}.

Obviously, increasing the number of time-series improves correspondence between
theory and simulations. However, once we approach a critical value of stochasticity,
at which a particular cumulant becomes divergent, computationally it is possible to
observe only the general trend towards theoretical result. Once $\sigma^2>J$, variance and
higher order cumulants no longer exist and we only observe relaxation of the mean,
in agreement with theory.

\begin{figure*}[!htbp]
\centering
\begin{tabular}{ccc}
\includegraphics[width = \myFigureWidth \textwidth]{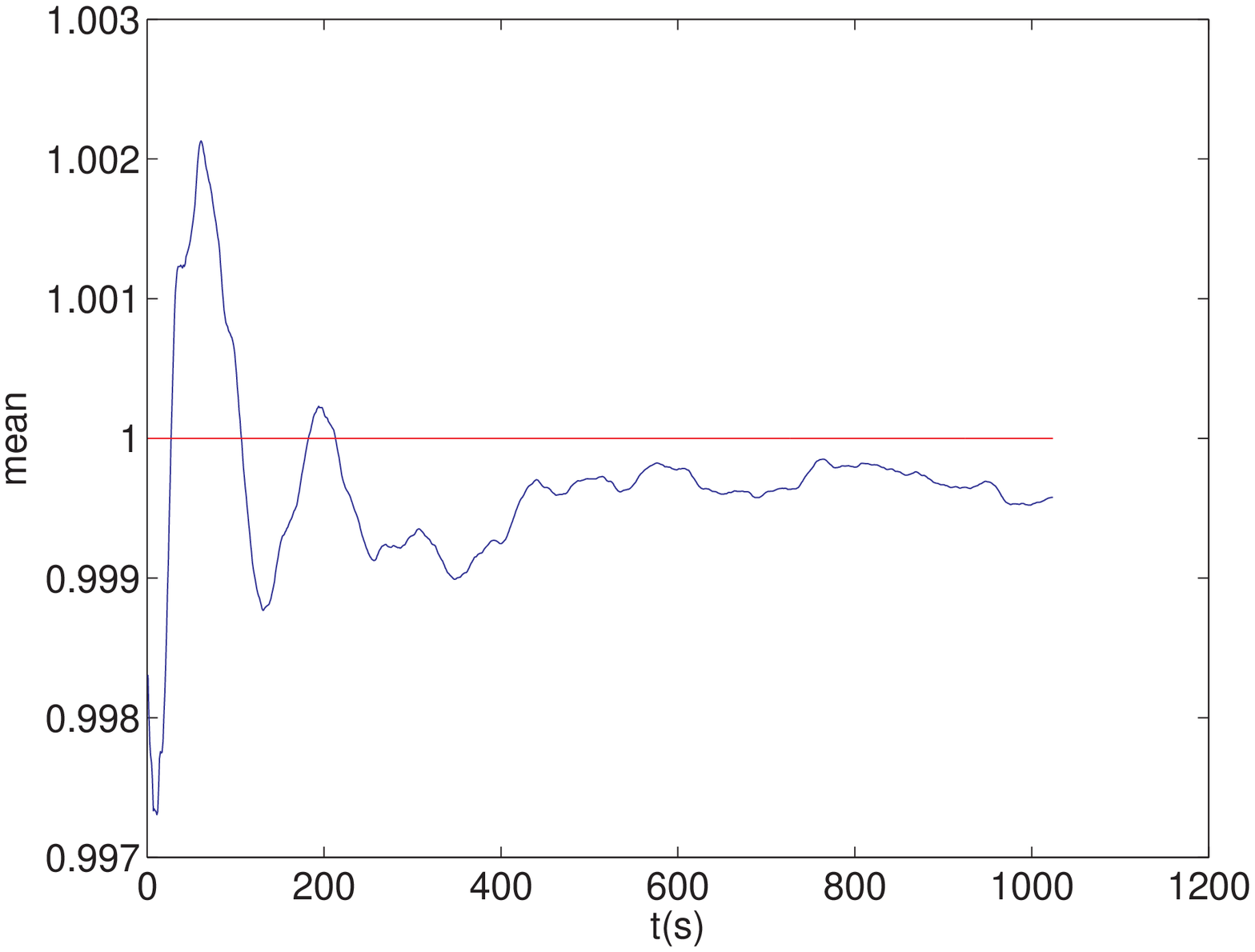} &
\includegraphics[width = \myFigureWidth \textwidth]{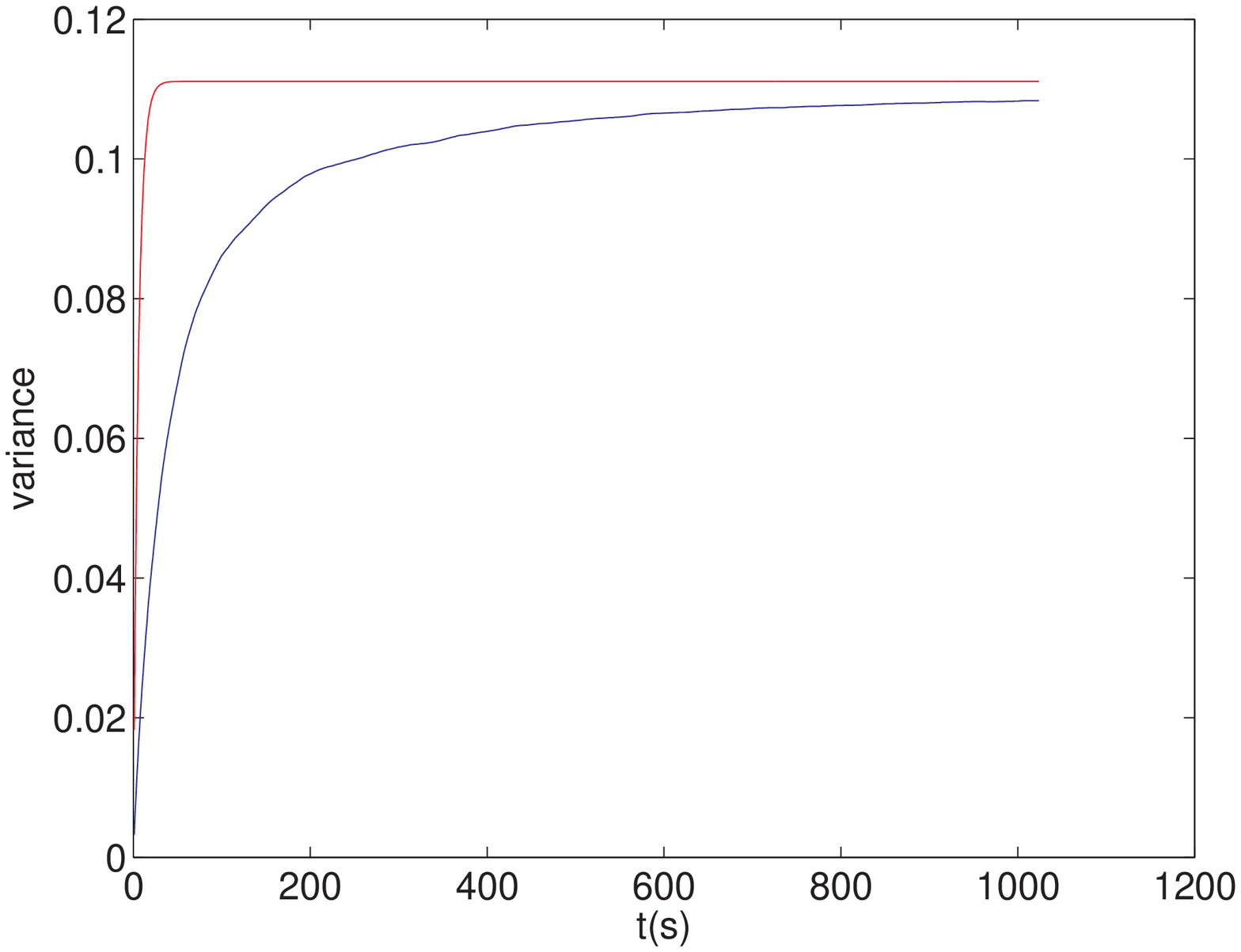} &
\includegraphics[width = \myFigureWidth \textwidth]{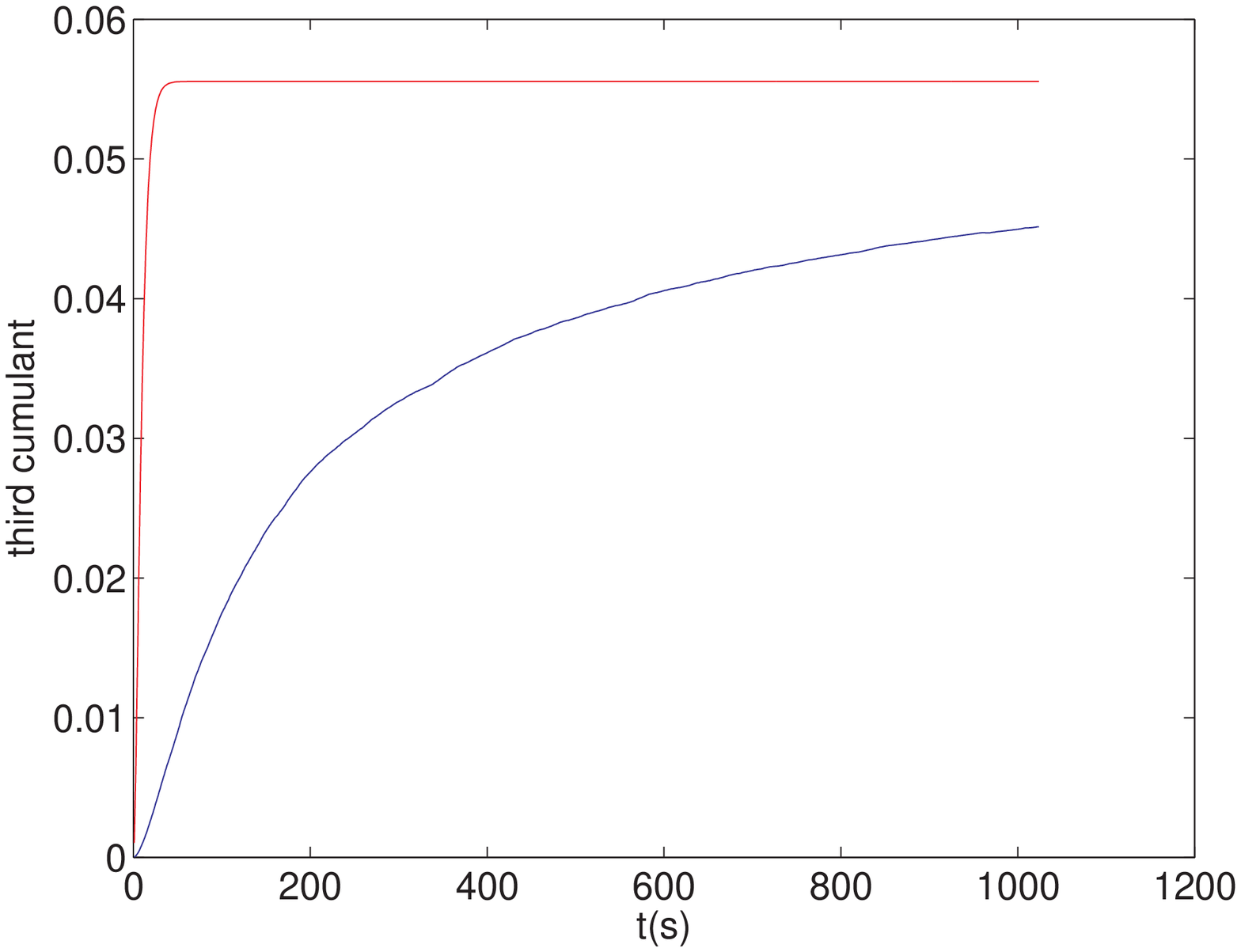} \\
\end{tabular}
\caption{Saturation of the mean, variance, and third cumulant vis-$\mathrm{\grave{a}}$-vis theoretical result, $\langle x\rangle = 1$, (\ref{correlationPaperKappa2AtZero}) and (\ref{correlationPaperKappa3}) for $x(0)=1$ for $10^4$ time series respectively, here $J=10^{-1}$ and $\sigma^2=10^{-2}$, as in the top plot of Fig. (\ref{meanVarianceSkewTrend}) -- "longitudinal" averaging.}
\label{meanVarianceSkewTrendLong}
\end{figure*}

This study of cumulant relaxation can be labeled "transverse" as we take a data point from each path at every time step. Alternatively, one can conduct a "longitudinal" study, where cumulants are evaluated along each path and the result is averaged over paths at every time step. For cumulants, longitudinal averaging is more computationally intensive. For comparison, in Fig. \ref{meanVarianceSkewTrendLong} we present a longitudinal plot with the parameters of the top plot of Fig. \ref{meanVarianceSkewTrend}. Clearly, it takes much longer to approach the theoretical values longitudinally. However the implication is that transverse and longitudinal results are equivalent. Conversely, when studying relaxation of the entire distribution to the steady state  and distribution of relaxation times (see next Section), longitudinal studies are more computation-friendly.

\section{Relaxation of the entire distribution and distribution of relaxation times \label{DistributionOfRelaxationTimes}}

\begin{figure}[!htbp]
\centering
\begin{tabular}{cc}
\includegraphics[width = \myFigureWidth \textwidth]{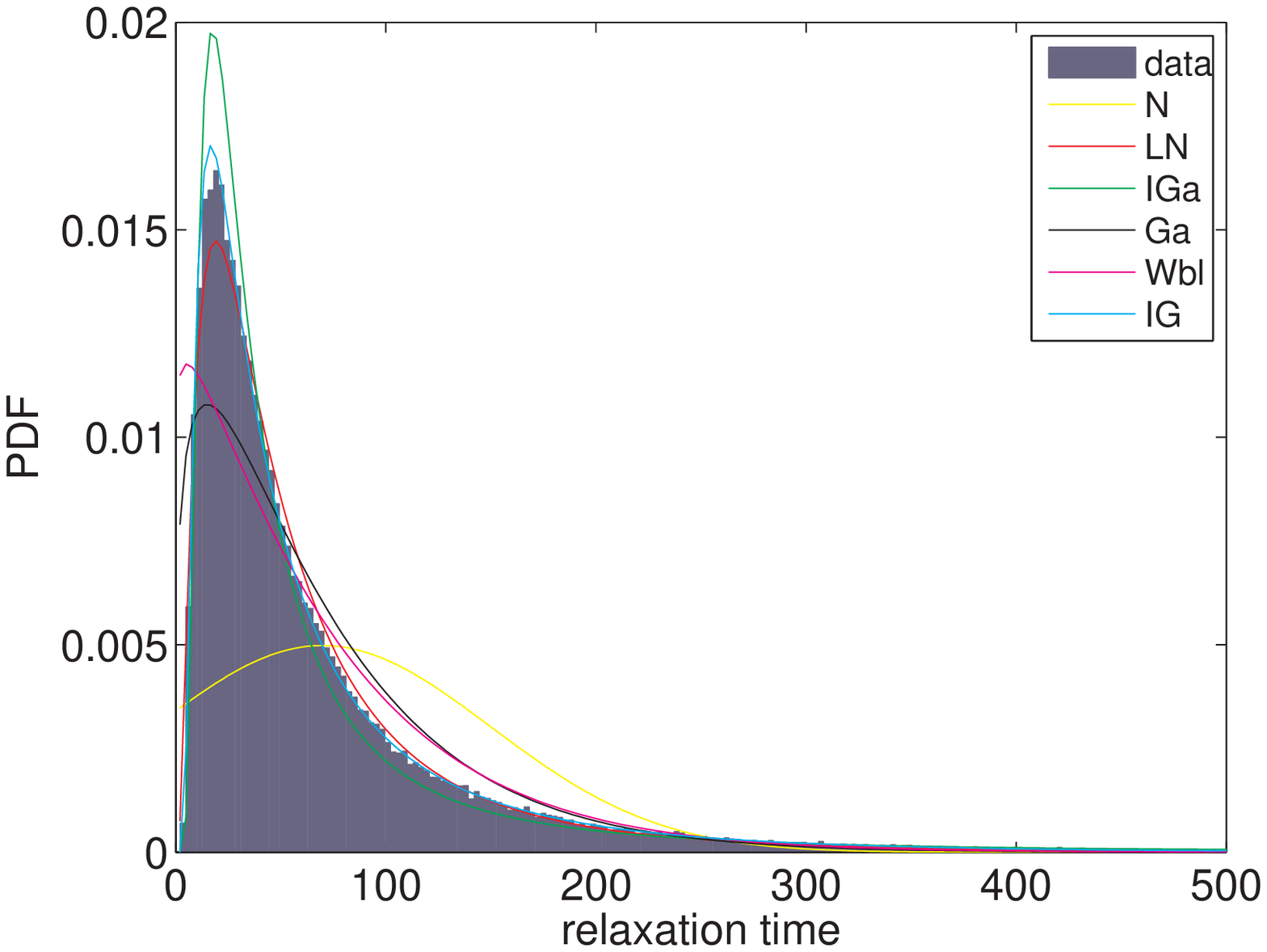} &
\includegraphics[width = \myFigureWidth \textwidth]{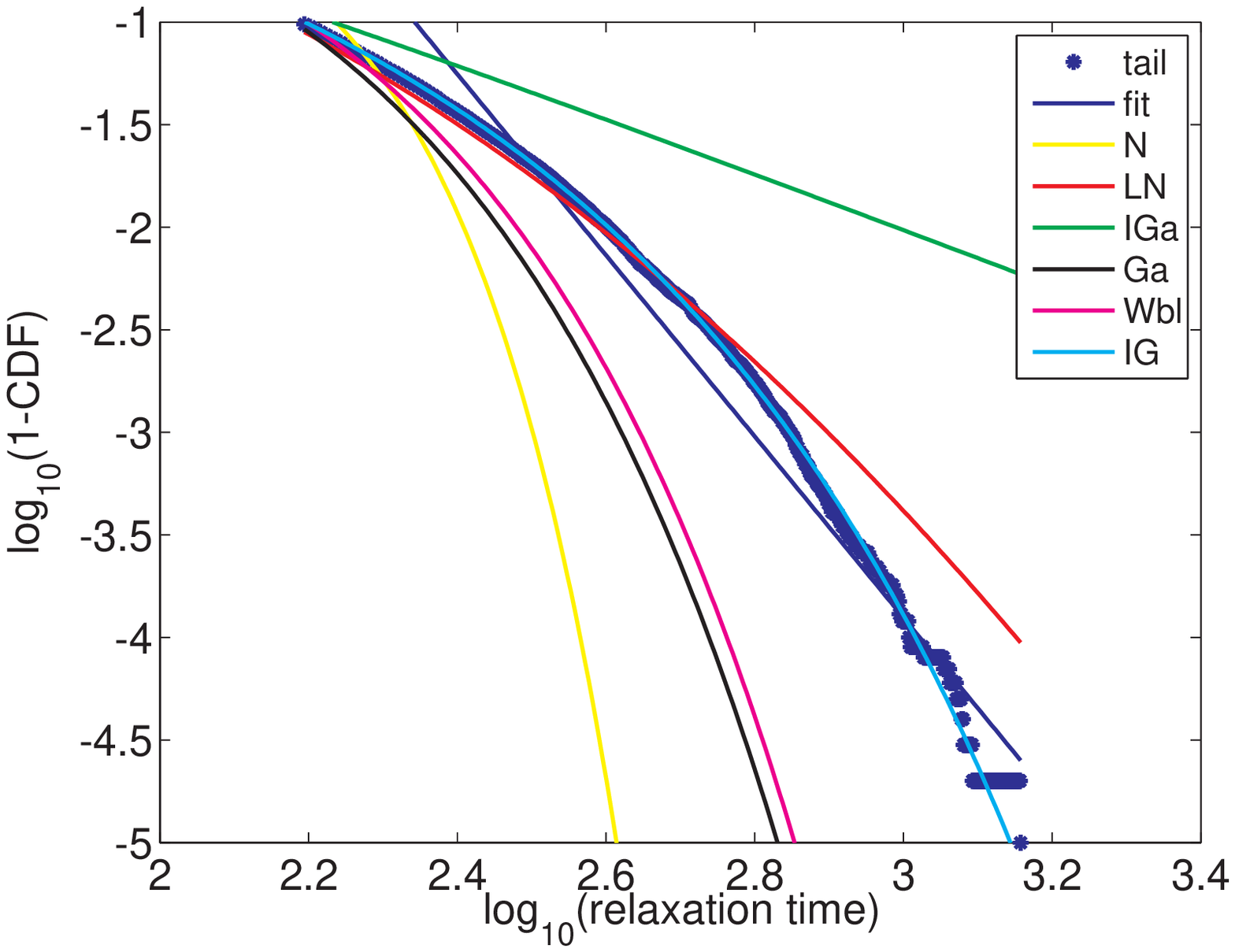} \\
\includegraphics[width = \myFigureWidth \textwidth]{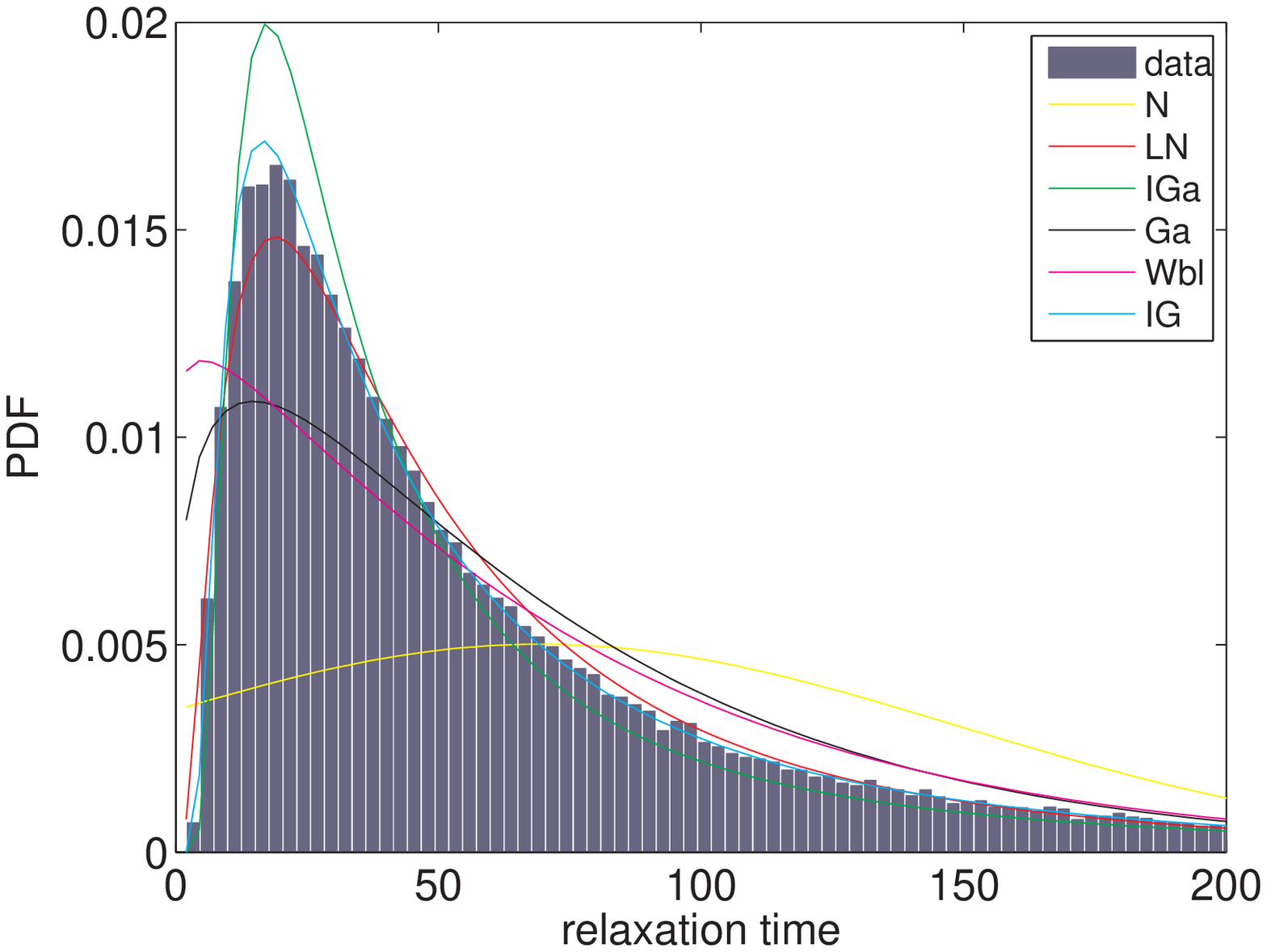} &
\includegraphics[width = \myFigureWidth \textwidth]{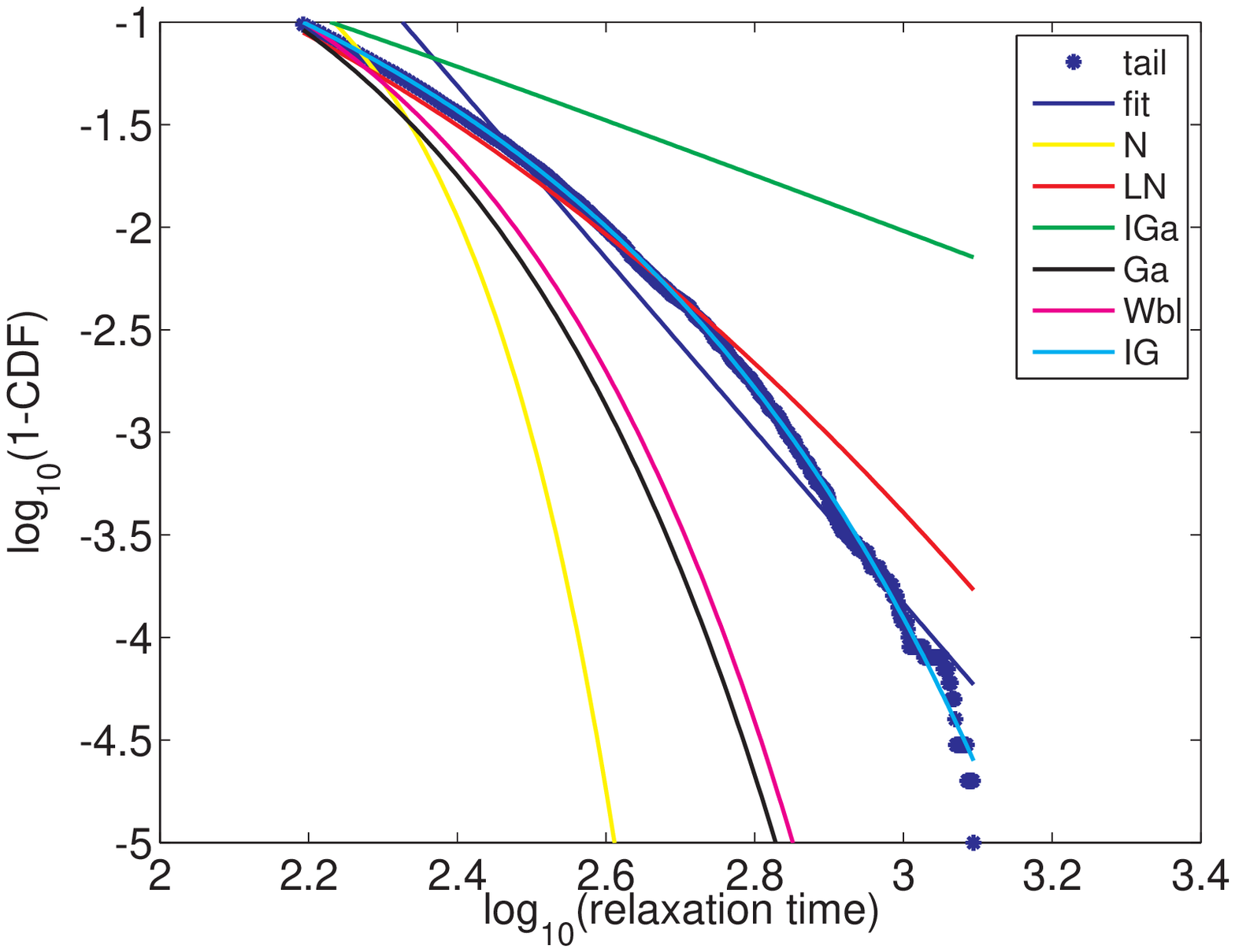} \\
\includegraphics[width = \myFigureWidth \textwidth]{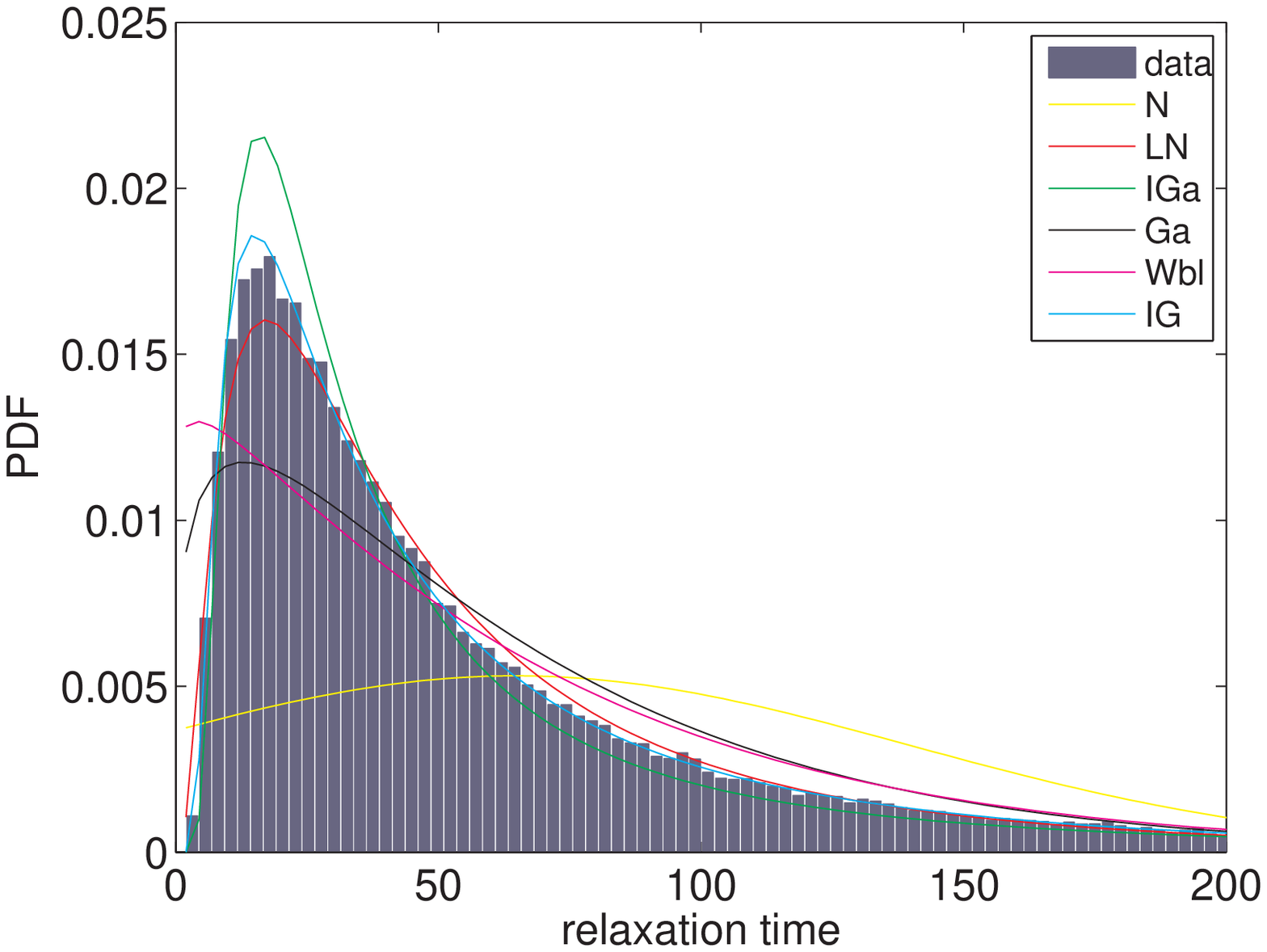} &
\includegraphics[width = \myFigureWidth \textwidth]{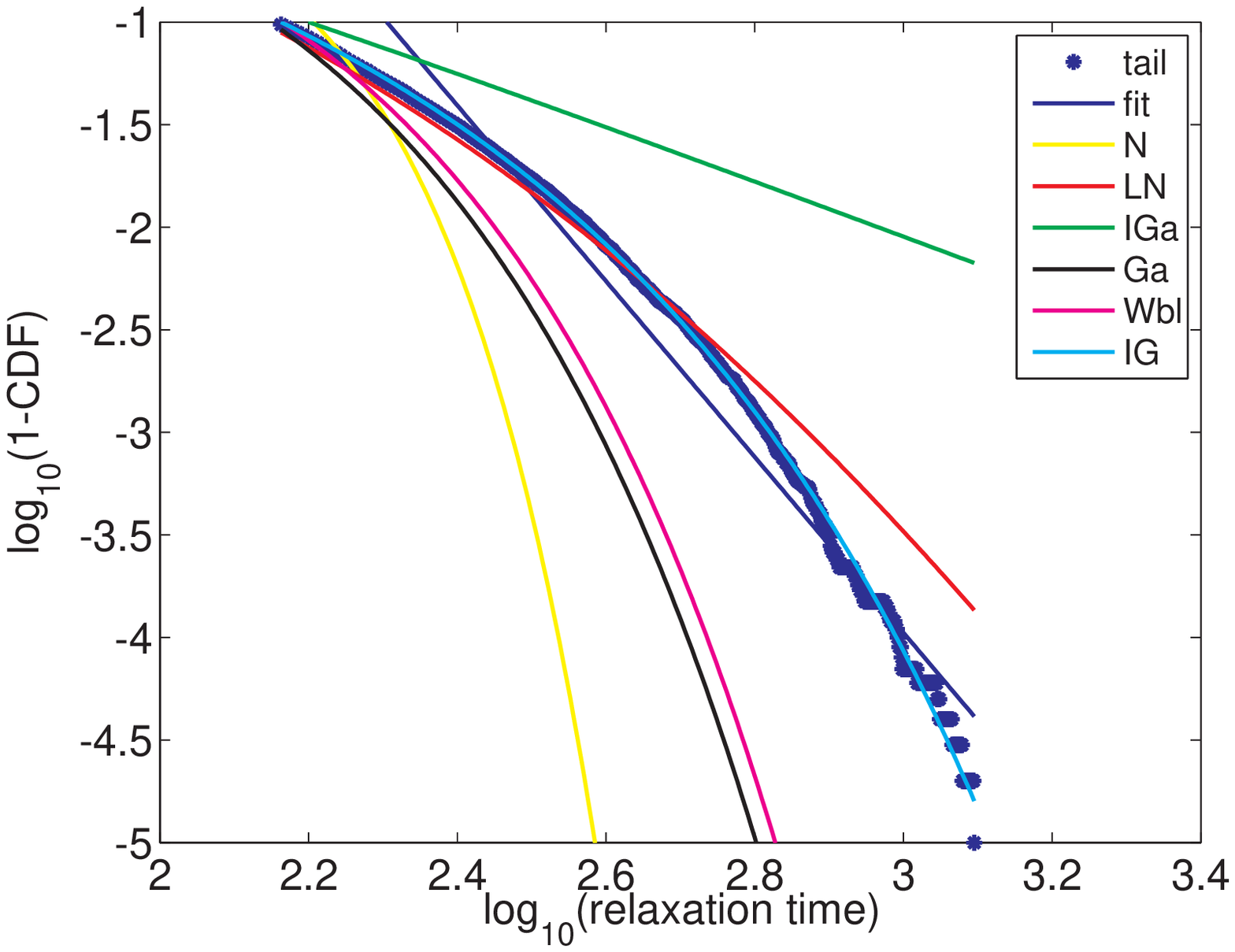} \\
\includegraphics[width = \myFigureWidth \textwidth]{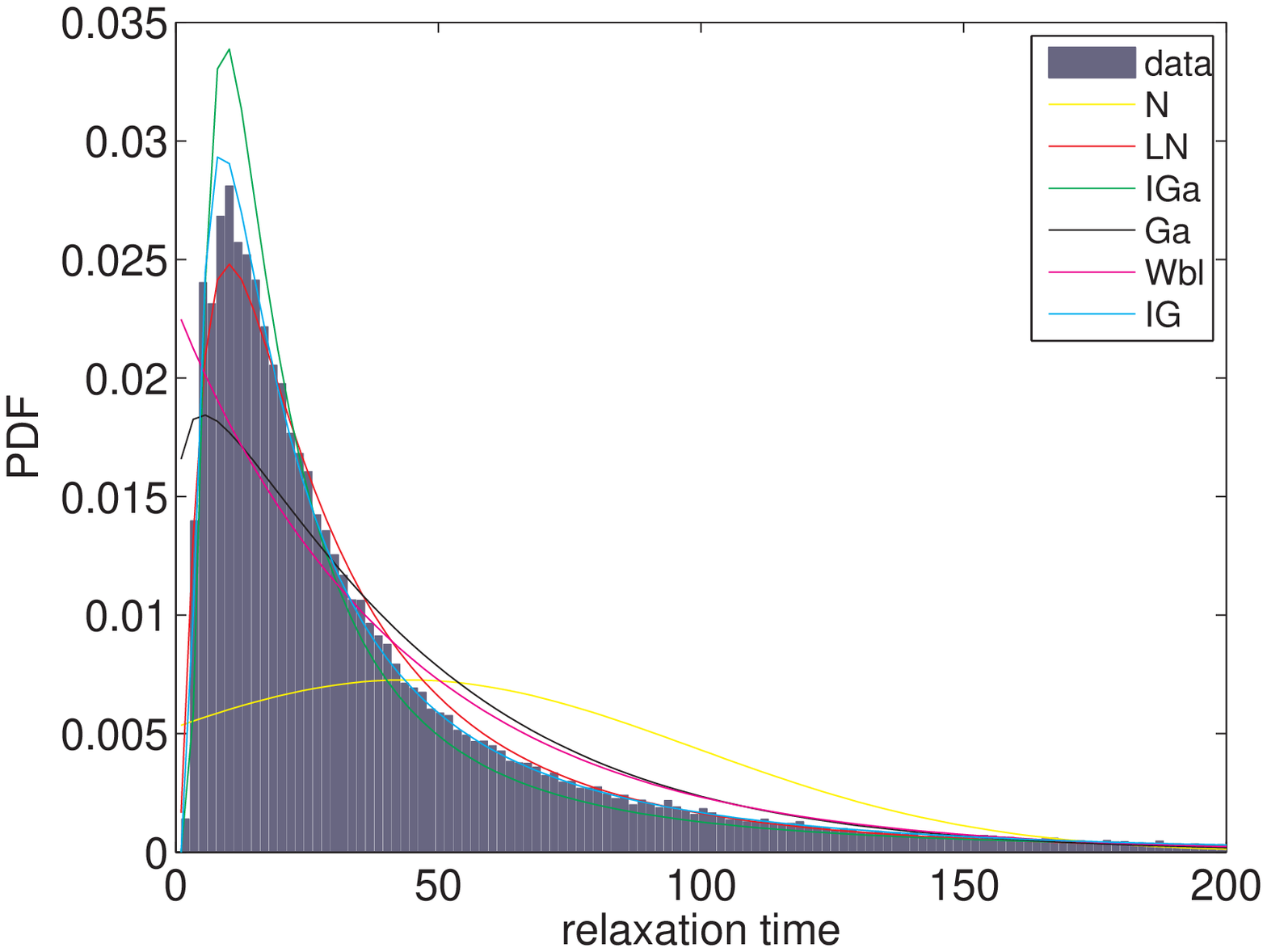} &
\includegraphics[width = \myFigureWidth \textwidth]{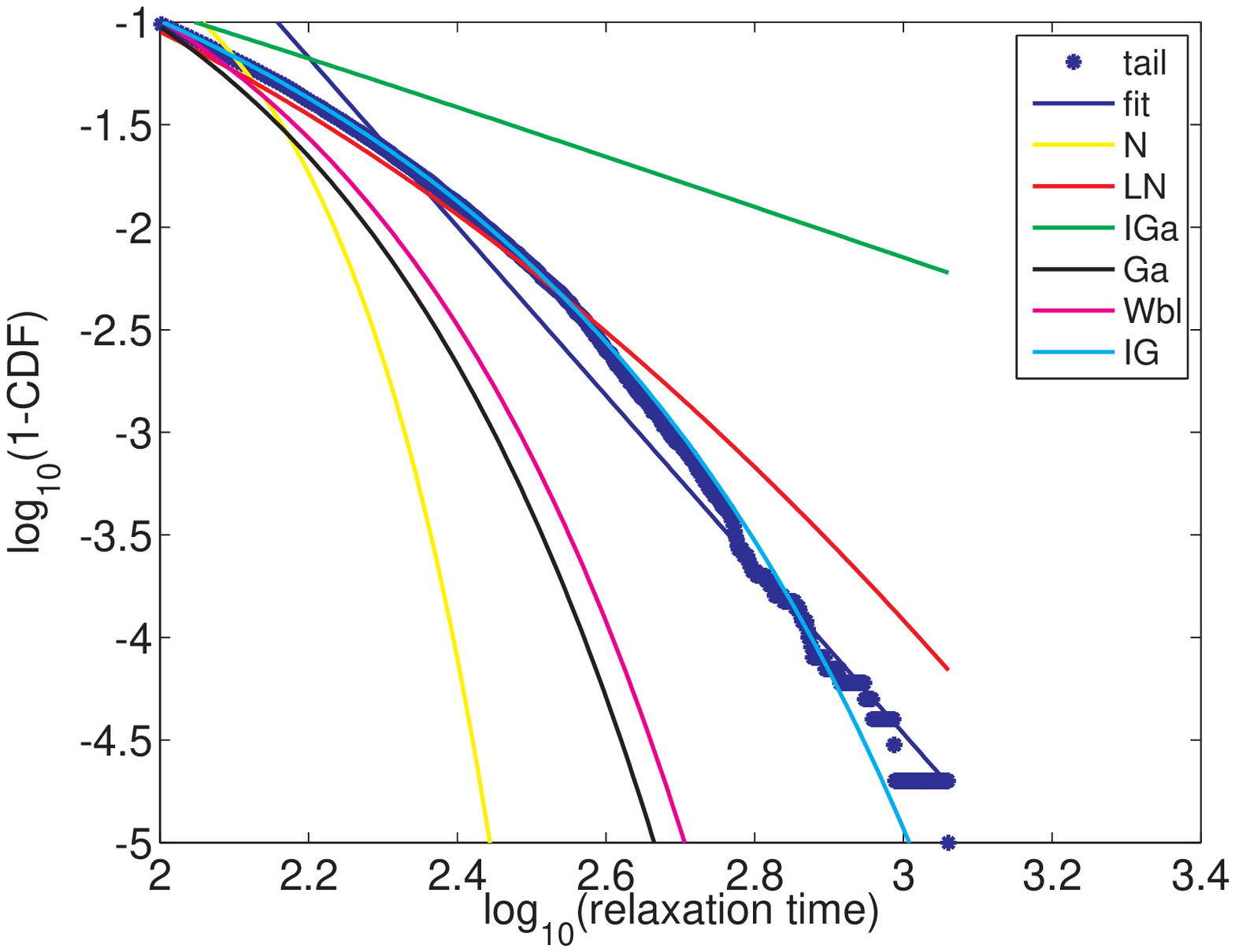} \\
\includegraphics[width = \myFigureWidth \textwidth]{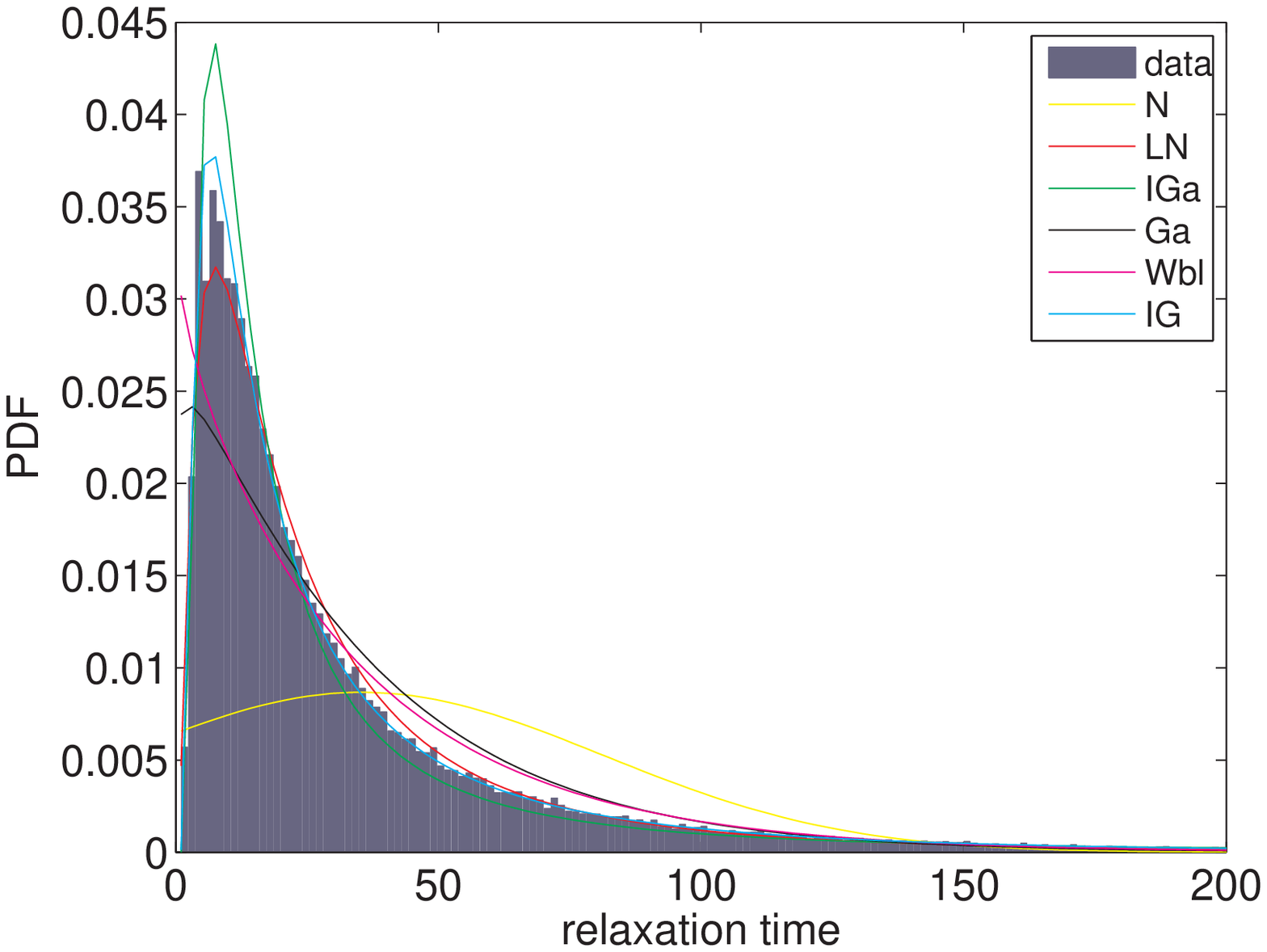} &
\includegraphics[width = \myFigureWidth \textwidth]{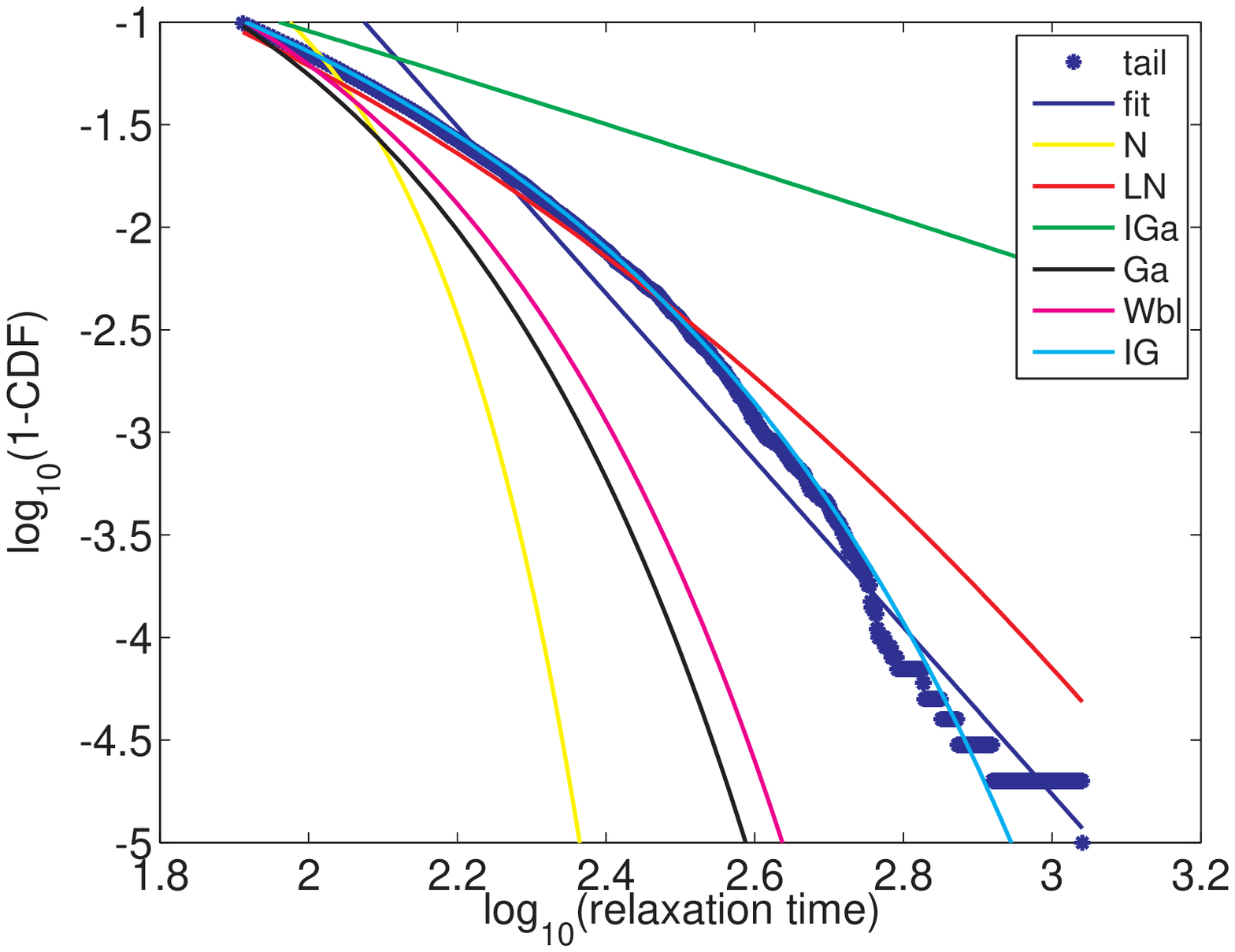} \\
\end{tabular}
\caption{From top to bottom, distribution fits, and tail fits for $J=10^{-1}$ and $\sigma^2 = 10^{-4}$, $1.12\times10^{-3}$, $1.27\times10^{-2}$, $0.89\times10^{-1}$ and $1.44\times10^{-1}$ respectively.}
\label{KStestFitTailfit}
\end{figure}

In view of diverging cumulants and their relaxations times studied in the previous Section, a question arises of establishing whether the entire distribution has relaxed to its steady-state. Accordingly, here we conduct "longitudinal" studies of such relaxation. Namely, for each paths discussed in the previous section we use the same small value of parameter in the Kolmogorov-Smirnov (KS) test to find the time at which the distribution of the time series generated by (\ref{correlationPaperItoProcess}) approaches the distribution given by (\ref{correlationPaperSteadyStateDistribution}). In this manner, we generate $10^{5}$ relaxation times to study their distribution as a function of $J$ and $\sigma^2$. We argue that relaxation times are distributed as IG. In order to ascertain the latter, we fit the relaxation-time distribution with six candidate distributions: Normal (N), Lognormal (LN), IGa, Gamma (Ga), Weibull (Wbl) and IG. Additionally, we fit the log-log tail of the distribution with a straight line to see if the tail may be power-law. The parameters of the distributions are obtained using Maximum Likelihood Estimation (MLE) and comparison of fitted distributions with the one obtained numerically is done via KS test. The results are presented in Fig. \ref{KStestFitTailfit} and Table \ref{TableMLEKS}. 

\begin{table*}[!htbp]
\centering
\caption{MLE fitted parameters and KS test values for $J=10^{-1};$ from left to right, the same values of $\sigma^2$ as top to bottom in Fig. \ref{KStestFitTailfit}.}
\label{tableMLEKS}
\begin{tabular}{lclclclclc}
\hline
\multicolumn{2}{c}{$\sigma^2 = 10^{-4}$} & \multicolumn{2}{c}{$\sigma^2 = 1.12\times10^{-3}$} & \multicolumn{2}{c}{$\sigma^2 = 1.27\times10^{-2}$} & \multicolumn{2}{c}{$\sigma^2 = 0.89\times10^{-1}$} & \multicolumn{2}{c}{$\sigma^2 = 1.44\times10^{-1}$} \\
\hline
\multicolumn{1}{c}{MLE} & KS & \multicolumn{1}{c}{MLE} & KS & \multicolumn{1}{c}{MLE} & KS & \multicolumn{1}{c}{MLE} & KS & \multicolumn{1}{c}{MLE} & KS \\
\hline
N(69.87, 80.07) & 0.206 & N(69.42, 79.58) & 0.206 & N(64.65, 75.05) & 0.209 & N(43.88, 54.87) & 0.224 & N(35.36, 46.00) & 0.233 \\
LN(3.80, 0.93) & 0.018 & LN(3.80, 0.93) & 0.018 & LN(3.72, 0.94) & 0.018 & LN(3.28, 0.99) & 0.018 & LN(3.03, 1.02) & 0.019 \\
IGa(1.39, 41.60) & 0.046 & IGa(1.39, 41.30) & 0.046 & IGa(1.37, 37.47) & 0.046 & IGa(1.25, 21.01) & 0.049 & IGa(1.19, 15.30) & 0.050 \\
Ga(1.27, 54.99) & 0.080 & Ga(1.27, 54.70) & 0.080 & Ga(1.25, 51.74) & 0.082 & Ga(1.13, 38.82) & 0.086 & Ga(1.07, 33.00) & 0.088 \\
Wbl(71.88, 1.06) & 0.070 & Wbl(71.41, 1.06) & 0.070 & Wbl(66.25, 1.05) & 0.071 & Wbl(43.82, 1.00) & 0.071 & Wbl(34.77, 0.97) & 0.074 \\
IG(69.87, 52.63) & 0.008 & IG(69.42, 52.23) & 0.008 & IG(64.65, 47.52) & 0.008 & IG(43.88, 27.40) & 0.013 & IG(35.36, 20.24) & 0.015 \\
\hline
\end{tabular}
\label{TableMLEKS}
\end{table*}

\begin{figure*}[!htbp]
\centering
\begin{tabular}{ccc}
\includegraphics[width = \myFigureWidth \textwidth]{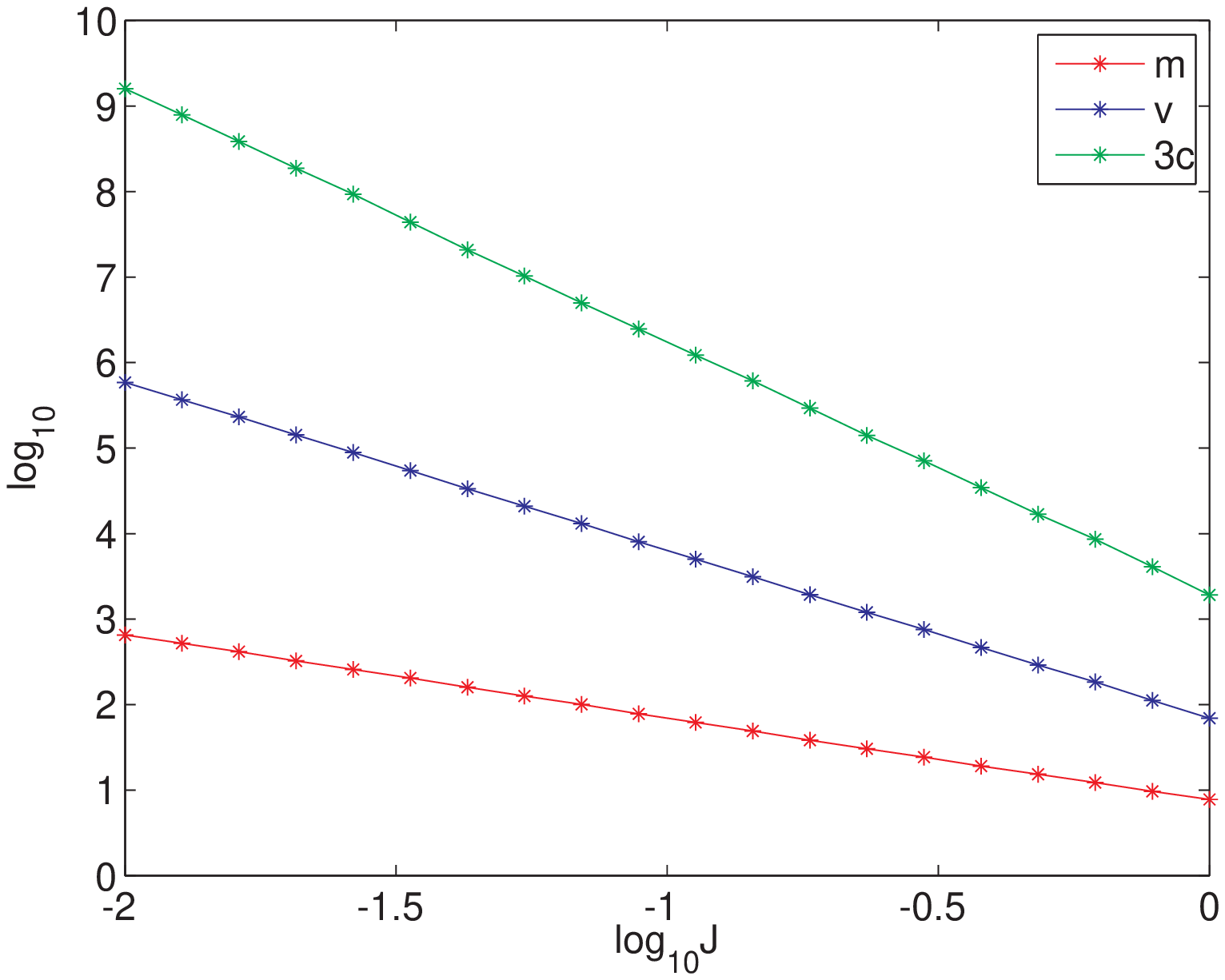} &
\includegraphics[width = \myFigureWidth \textwidth]{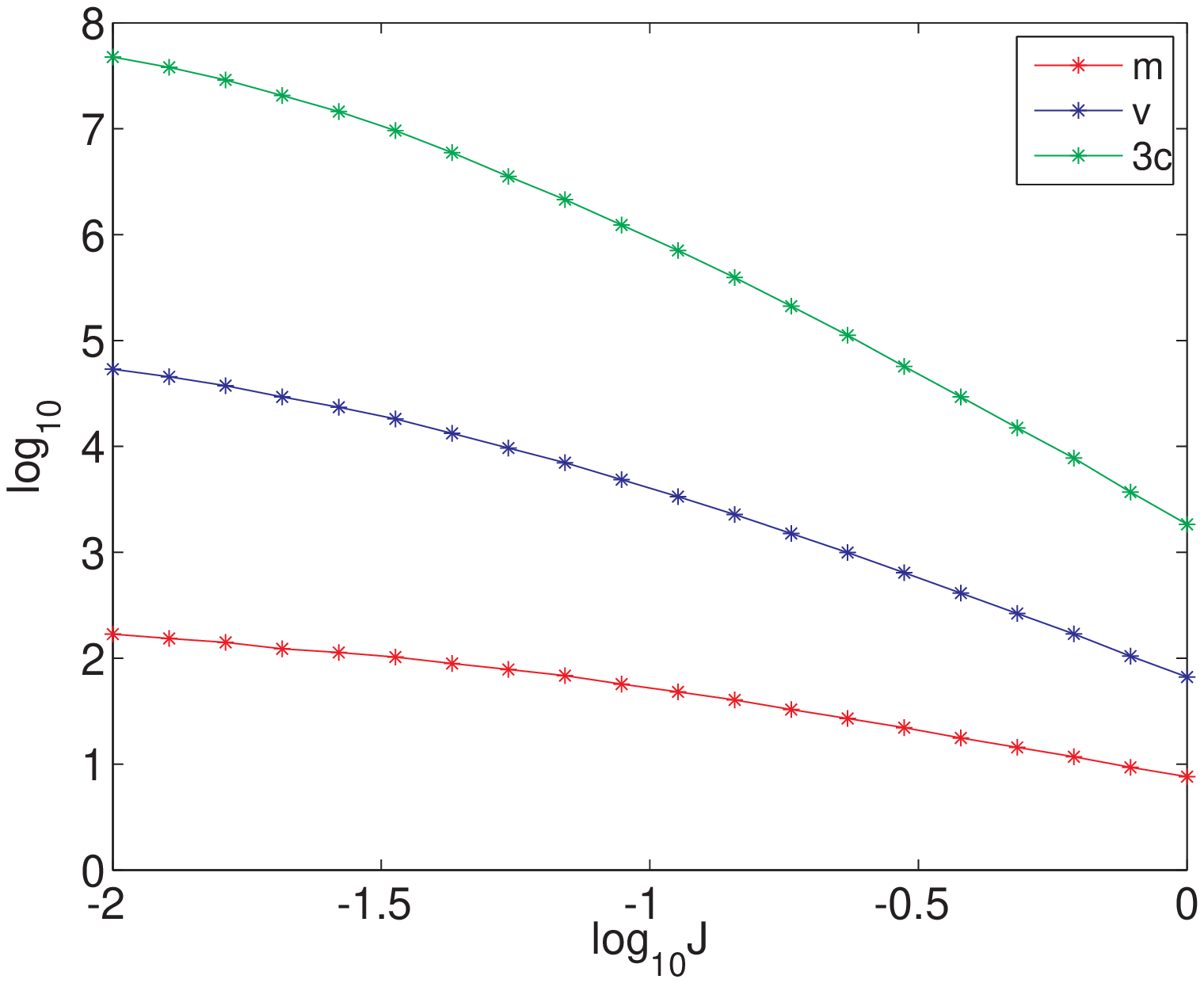} &
\includegraphics[width = \myFigureWidth \textwidth]{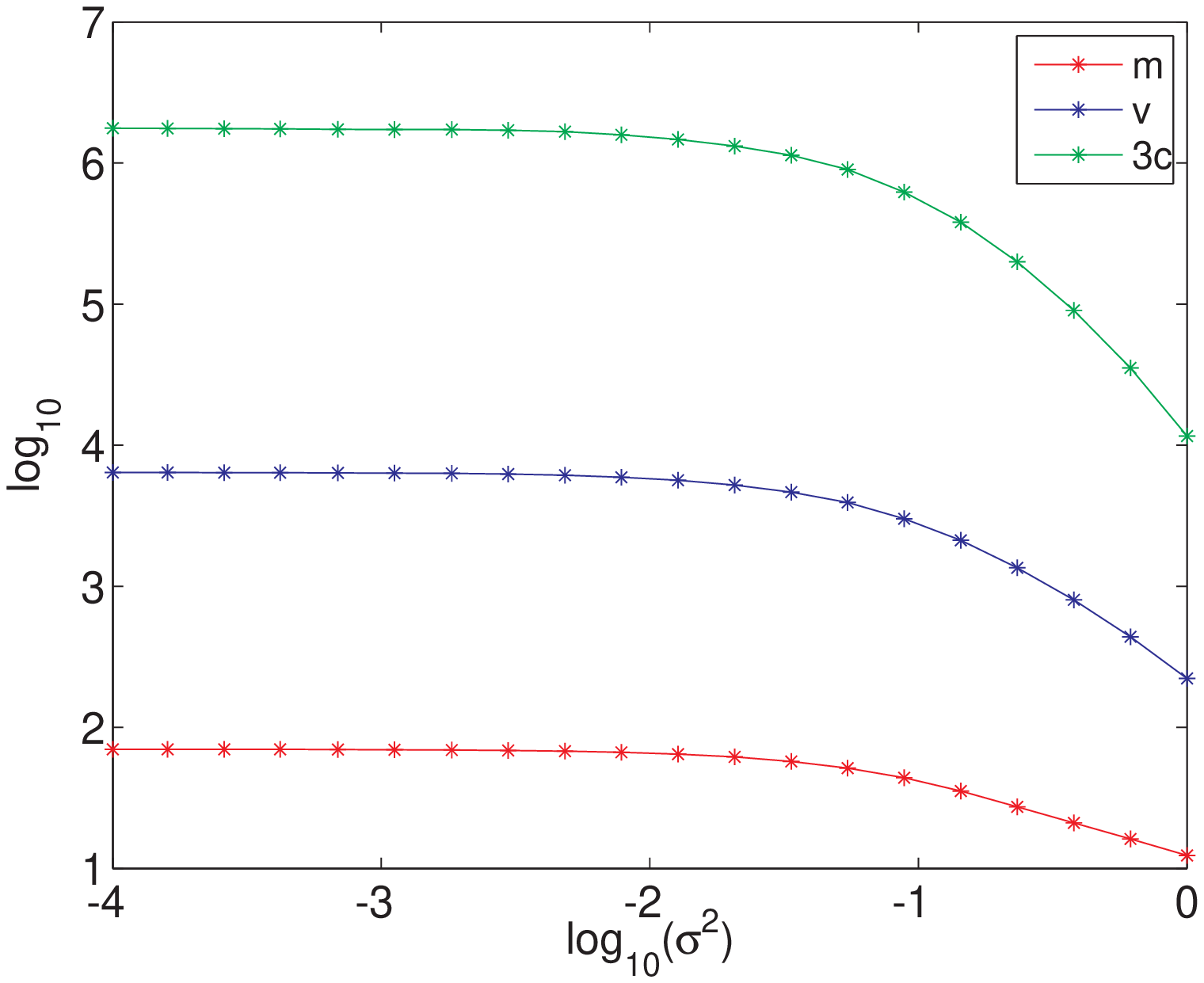} \\
\end{tabular}
\caption{Dependence of the mean, variance and third cumulant of the relaxation time distribution on $J$ for $\sigma^2=10^{-3}$ (left) and $\sigma^2=5\times10^{-2}$ (middle) and on $\sigma^2$ for $J=10^{-1}$ (right); $J$ varies between $10^{-2}$ and $1$, $\sigma^2$ varies between $10^{-4}$ and $1$.}
\label{LogMeanVarianceSkewLogJ}
\end{figure*}

Clearly, IG constitutes the best fit and can be presented in the following form:
\begin{equation}
IG(aJ^{-1}, bJ^{-1}; x) = \sqrt{\frac{b}{2 \pi  J x^3}} \exp{\left[-\frac{b J(x-a J^{-1})^2}{2 a^2 x}\right]}
\label{IG}
\end{equation}
where $a$ and $b$ are constants and the distribution does not depend on $\sigma$. 
\footnote{While not presented here, the distribution of relaxation times for OU process is similarly best fitted by an IG dependent only on $J$.}
To further verify (\ref{IG}), we observe that $n$'th cumulant of this distribution scales as $\kappa _n\propto J^{-n}$. In Fig. \ref{LogMeanVarianceSkewLogJ}, we plot, on the log-log scale, the first three cumulants as a function of $J$ and $\sigma^2$ respectively. Obviously, with the the exception of $\sigma^2 > J$, where simulations become unreliable, it lends support to our conclusions vis-a-vis IG (\ref{IG}).

While we presently do not have a first-principles model for explaining the IG distribution for relaxation times, we speculate why it has the necessary properties \cite{chikkara1988inverse, seshadri1998inverse}:

\begin{itemize}
\item {The approach to the distribution has to be controlled by a single time scale $J^{-1}$ for any $\sigma^2 $ since it is the case when $\sigma^2 > J$ for IGa. \footnote{For OU, of course, $J^{-1}$ is the only relaxation and correlation scale.}}
\item {On physical grounds, it is clear that for relaxation times the distribution of the sample mean should have the same distribution as the distribution from which the sample is taken. This, of course, is also the property of the IG distribution. \footnote{Of course, as the sample size increases, the IG of the mean tends to the normal distribution, in agreement with central limit theorem.}}
\item {Time to achieve the steady-state distribution can be conjectured to be the first passage time in the distribution space, where it scales as $\propto J^{-1}$.}
\end{itemize}
\section{Conclusions \label{Conclusions}}

We conducted a study which examined in great detail relaxation times towards the fat-tailed, steady-state distribution of a stochastic process. Specifically, we examined the relaxation times of the cumulants of the IGa process, whose steady state is characterized by power-law tails. We found that, as stochasticity rises, successive lower cumulants diverge, as do their relaxation times. These divergencies are controlled by the inverse eigenvalues of the Fokker-Planck eigenvalue problem. We also found that the distribution of the relaxation times is best approximated by an IG distribution with a single time scale.

The implications of our findings may be multifaceted. We know for instance that the IGa process describes the mean-field limit of the BM economic network model \cite{bouchaud2000wealth}, while the more general GIGa process describes a partially connected network and stock market volatility \cite{ma2013distribution,ma2014model}. While perhaps unrelated, it should be noted that the importance of multiple time scales and relaxation phenomena has been widely recognized in financial markets and economic models -- see \cite{borland2007multi-timescale} and \cite{yakovenko2009statistical} and references therein.

In future work we would like to complete the transverse relaxation study of the distribution in relation to the longitudinal one. It would be interesting to compare those vis-a-vis the relaxation studies of the wealth distribution for and individual versus that of the entire group of participants in BM economic exchange. We would also like to extend our relaxation studies to those of stock returns and volatility.

\bibliography{mybib}

\end{document}